# Magnetic on-surface assemblies predicted from a pious computational method

Daniel M. Packwood

Institute for Integrated Cell-Material Sciences (iCeMS), Kyoto University, Kyoto 606-8502, Japan

**Abstract**

Molecular self-assembly will not become a routine method for building nanomaterials unless our ability to predict the outcome of this process is dramatically improved. Even then, reliable strategies for realizing molecular assemblies with novel properties are required for building nanomaterials for specific device applications. In this paper, I simulate the self-assembly of metal phthalocyanine derivatives adsorbed to gold(111) surfaces using a detailed statistical mechanical model and a new computational method based upon genetic algorithms and Markov chain Monte Carlo (MCMC). This method yields predictions that are not only are superior to those of ordinary MCMC but also show good agreement with experimental results. Crucially, it is predicted that molecular assemblies displaying locally disordered magnetic moments – and having potential applications as non-Gaussian noise sources for device applications – can be realized by the simple strategy of introducing asymmetry into the phthalocyanine ligands.

## 1. Introduction

Nearly 160 years have passed since Samuel Butler published a keen observation on the nature of technological progress: "...so a diminution in the size of machines has often attended their development and progress" [1]. Since then, device miniaturization and integration has advanced far beyond what Butler could have imagined, and it has become easy to take his observation and similar ones such as Moore's law for granted. Yet this diminution cannot continue indefinitely; eventually it will be limited by the extent to which device components can be built bottom-up from molecule building blocks.

How does one go about constructing device components from individual molecules? For many materials chemists, the answer is molecular self-assembly. Molecular self-assembly is a phenomenon in which isolated molecules spontaneously assemble into supramolecular structures. The molecular self-assembly phenomenon has garnered an enormous literature over the last few decades, which has been reviewed in detail elsewhere [2, 3, 4, 5]. What has not been emphasized, however, is the fact that materials chemists make a huge conceptual leap when presenting molecular self-assembly as a way of constructing device components. This leap occurs when molecular self-assembly – a natural phenomenon – is elevated to the status of a scientific methodology. As a result of this leap, two serious difficulties have not received sufficient attention.

The first difficulty is a consequence of the spontaneous nature of molecular self-assembly. The outcome of the molecular self-assembly process is predetermined by parameters ('assembly parameters') such as the structure of the molecules and the temperature. These assembly parameters control how the molecules interact with each other and which types of assemblies are favored at thermodynamic equilibrium. These assembly parameters are selected by the scientist before the self-assembly process starts, and once the process starts its outcome cannot be affected. As a methodology, molecular self-assembly therefore has an unappealing trial-and-error quality, in which the assembly parameters are selected haphazardly and tested in the hopes of obtaining the desired outcome. The situation becomes even less appealing when we consider the case of on-surface molecular assemblies, for which scanning tunneling microscopy (STM) is the standard characterization technique. Even for a single instance of assembly parameters, STM can demand months of work to obtain clear images of the resulting assembly. Years of work may be required if multiple instances of assembly parameters are to be tested. In order to enable molecular self-assembly as an effective methodology, significant breakthroughs in the our ability to characterize (or predict) the outcome of molecular self-assembly processes are utterly essential.

The second difficulty becomes clear when we consider the physical requirements for a molecular assembly to serve as a component of a device. As a component of an electronic device in particular, the molecular assembly should possess properties such as electrical conductivity or magnetic ordering [5, 6]. Such properties arise from the collective interaction between molecules and pertain to the assembly as a whole. Yet most molecular assemblies are nothing more than aggregates of molecules, lacking any remarkable collective properties. It is no coincidence that many prominent applications of molecular assemblies have been in the fields of biological chemistry [7] and host-guest chemistry [8], where assembly shapes, rather than collective physical properties, are decisive. Nonetheless, for molecular self-assembly to be useful as a method for creating components for devices, strategies for realizing collective physical properties are required.

While these two difficulties are rarely explicated in the literature, they have nonetheless influenced recent research trends, particularly for the case of on-surface molecular assemblies. In response to the first difficulty, much attention has shifted towards computational methods as a means of predicting on-surface assemblies without incurring the enormous costs of STM experiments [9]. Machine learning techniques have enabled remarkable progress in this area, and it is now possible to predict molecule adsorption conformations [10, 11], on-surface assemblies [12 - 14] and molecular monolayers [15 - 18] with density functional theory (DFT)-levels of precision within days of computation (compared to months for the case of some STM experiments). It is no longer fanciful to imagine a future where computers guide scientists towards the creation of desired assemblies. Yet much work is required before such a future can be realized. In particular, the predictive components of these computational methods, be they Bayesian optimization or Markov chain Monte Carlo (MCMC), are either intractable when large numbers of molecules are present on the surface or when applied at low molecule surface coverage.

As far as efforts to address the second difficulty are concerned, one strategy holds great promise: the use of porphyrin-like metal complex molecules as a means towards molecular assemblies with magnetic order [19 - 21]. One of the most compelling studies in this area was published by Girovsky *et al* in 2017 [22], who observed antiferromagnetic ordering in a self-assembled monolayer of FeFPc (iron fluoro-phthalocyanine) and MnHPc (manganese phthalocyanine) molecules adsorbed to a gold(111) (Au(111)) surface. The structures of FeFPc and MnHPc molecules are shown in Figure 1A and 1B, respectively. The antiferromagnetic ordering observed in this system results from a combination of two factors. The first factor is the presence of surface-mediated Ruderman-Kittle-Kasuya-Yoshida (RKKY) interactions between the unpaired electrons of the Mn and Fe ions of neighboring molecules. The second factor is the presence of strong hydrogen bonding between the FeFPc and MnHPc molecules, which drives the formation of a densely packed, tile-like assembly. The second factor is crucial, as previous research had found that the Kondo effect, which is expected to screen the RKKY interaction, is weakened as the density of molecules in the assembly increases [23]. The possibility of tuning these molecular assemblies by judicious structural alterations to the FeFPc and MnHPc ligands, and hence observing entirely new types of magnetic order, is an unprecedented opportunity in materials chemistry. To the best of my knowledge, this opportunity remains unexplored.

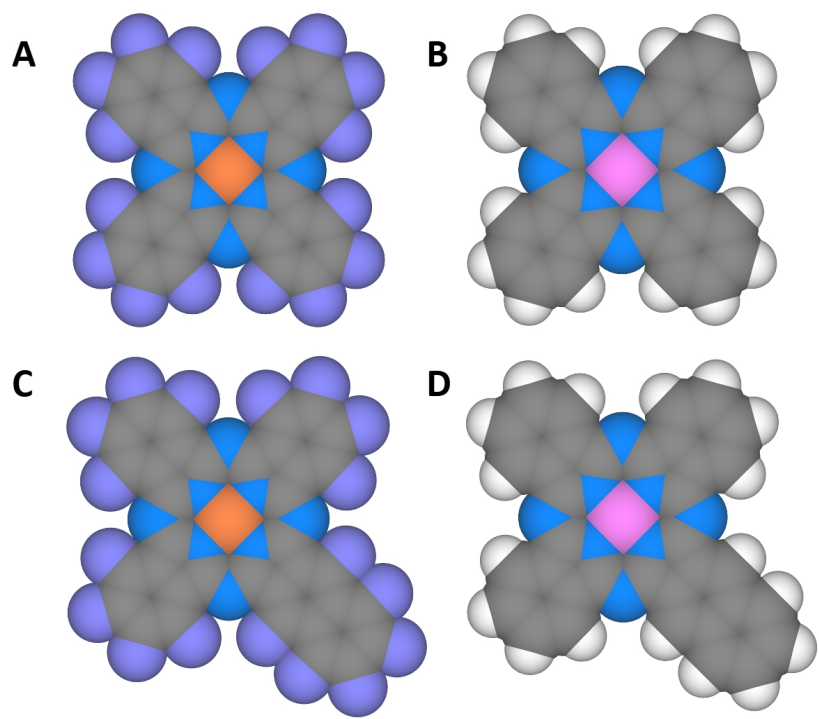

**Figure 1.** Molecular models of (A) symmetric iron fluoro-phthalocyanine (FeFPc) (B) symmetric manganese phthalocyanine (MnHPc), (C) asymmetric iron fluoro-phthalocyanine (AFeFPc), and (D) asymmetric manganese phthalocyanine (AMnHPc), as viewed top-down on the molecular plane. Grey = carbon, blue = nitrogen, purple-blue = fluorine, white = hydrogen, orange = iron, purple = manganese. All molecular models were drawn with the VESTA software [51].

In this paper, I explore this opportunity by way of a new computational method. This computational method can predict the outcome of the self-assembly of porphyrin-like metal complex molecules adsorbed to metal surfaces. It refines previous work [12], which embedded atomistic detail into a statstical mechanical model using machine learning, while incorporating an efficient new prediction methodology based on simulated annealing, genetic algorithms, and MCMC. I show that by replacing the ligands of the FeFPc and MnHPc molecules with asymmetric ones (Figure 1C, 1D), it is possible to induce the formation of a liquid-crystal-like molecular assembly, in contrast with the densely packed, crystal-like molecular assemblies described in previous work. Interestingly, these liquid crystal-like assemblies can display magnetic order while simultaneously possessing agitated spins which undergo continuous spin flips, which is suggestive of potential applications in spintronics devices or as non-Gaussian noise sources.

This paper is organized as follows. Section 2 describes the physical model on which the computational method is based, and Section 3 explains the new prediction methodology. Section 4 presents predictions for the self-assembly of FeFPc and MnHPc molecules on Au(111) surfaces, and compares them to predictions for their asymmetric analogues. The magnetic ordering of the resulting molecular assemblies is discussed in Section 5. Discussion and conclusions are left to Section 6.

**2. Ising-like model with embedded atomistic detail**

Consider a small number of FeFPc and MnHPc molecules freshly deposited onto a gold(111) (Au(111)) surface under vacuum. At first, the molecules would be adsorbed at random locations on the surface and oriented in random directions. The subsequent drive towards thermodynamic equilibrium occurs as the molecules adjust their positions and orientations in order to optimize their interactions with each other and the surface. In doing so the molecules would be expected to condense into an ordered molecular assembly. Our task is to predict these molecular assemblies using computational methods.

Fully atomistic simulations, whether they are performed from first principles or with classical force fields, are plainly unsuitable for such purposes. Not only are thousands of atoms involved, but the equilibrium state may require microsecond-exceeding simulation trajectories to access. Even putting these computational barriers aside, charge transfer effects between the surface and molecules means that the system cannot be modeled as a collection of balls and springs, which rules out approaches based on classical mechanics. In this study I therefore employ an alternative approach, in which the atomistic detail of the system is embedded into the parameters of a simple 'Ising-like' statistical mechanical model. The effectiveness of this approach has been demonstrated previously [12, 15].

This Ising-like model is illustrated in Figure 2A. A perfectly crystalline Au(111) slab of infinite size and frozen internal degrees of freedom is considered, on top of which a grid of evenly-spaced adsorption sites is placed. $n_1$ of these adsorption sites are occupied by an FeFPc molecule, and $n_2$ are occupied by an MnHPc molecule. By 'occupied', we mean that the center of these molecules (the metal ion) lies some specified distance directly above the adsorption site. All molecules of the same type have identical and planar conformations, as well as frozen internal degrees of freedom. At each adsorption site, each molecule may adopt one of a fixed number of orientations. For a given configuration $c$ of the molecules (that is, a specific assignment of adsorption sites and orientations to each molecule), the energy of the system is given by

$$\epsilon(c) = \sum_{i=1}^{n_1+n_2} u_i + \sum_{i \neq j} v_{ij}, \qquad (1)$$

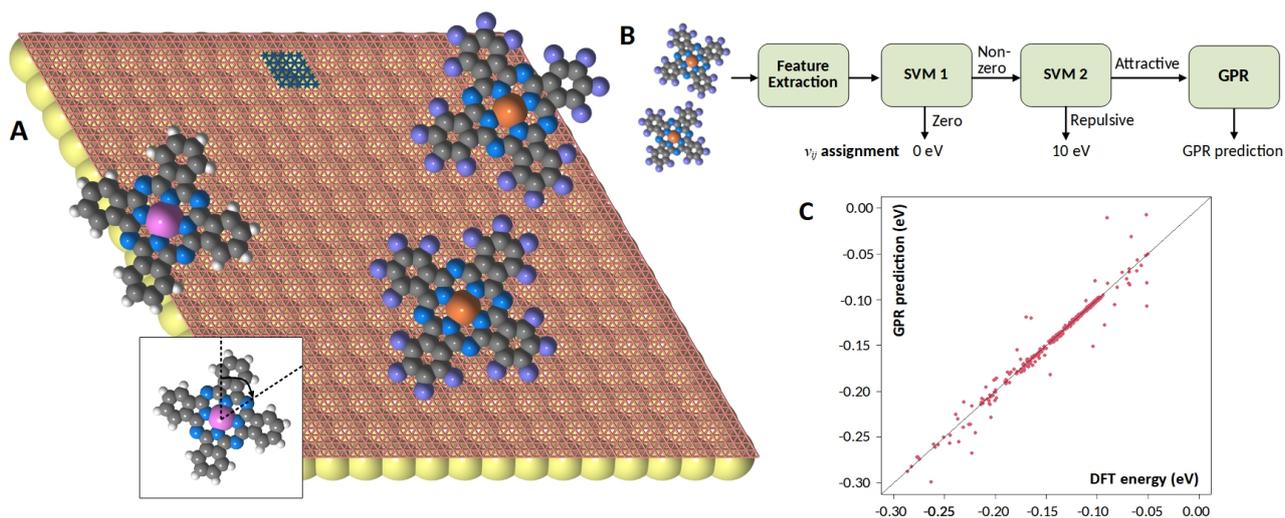

**Figure 2.** (A) Illustration of the Ising-like model used in this study. This model considers a grid of adsorption sites (orange mesh), upon which the molecules may adopt one of several orientations (orientation defined in the insert). The adsorption sites available at one gold(111) unit cell are indicated by in dark blue. Yellow spheres = gold atoms. (B) Protocol for assigning interaction energies to pairs of molecules. The protocol accepts a pair of molecules as input, extracts a feature vector from them, and then applies a series of machine-learned models (SVM1, SVM2, GPR) to assign the interaction energy. See text for details. (C) Comparison of interaction energies assigned by the protocol in (B) with test data acquired from density functional theory (DFT) calculations, for the case of symmetric molecules.

where $u_i$ is the interaction energy between the surface and molecule $i$, $v_{ij}$ is the energy of interaction between molecules $i$ and $j$, and the second sum runs over all unique pairs of molecules. The probability that configuration $c$ appears at thermodynamic equilibrium (the 'formation probability') is

$$p(c) = \frac{e^{-\epsilon(c)/k_B T}}{Z} \qquad (2)$$

where $Z$, $k_B$, and $T$ are the partition function, Boltzmann constant, and surface temperature, respectively. Expression (1) makes a number of simplifications which are discussed in Section 6.

The interaction energy parameters in equation (1) are set in a succession of steps *via* density functional theory (DFT) calculations and machine learning.

*Step 1. Set the molecule conformations and adsorption heights.* The conformations of the FeFPc and MnHPc molecules are determined by relaxing their structures in vacuum. The molecules are then placed at various distances above an unrelaxed Au(111) slab, and for each distance a single-point energy calculation is performed *via* DFT. In this manner, the distance at which the energy of the system is minimized (the adsorption height) is determined. The adsorption heights for FeFPc and MnHPc were found to be 3.37 A and 3.29 A, respectively.

The grid of adsorption sites and permitted orientations for the molecules are also set in Step 1. In this work, the gird was formed by dividing the Au(111) lattice vectors into five evenly-spaced points, resulting in a grid of 25 adsorption sites for each Au(111) cell (a 15 x 15 unit cell segment of the grid is shown shown Figure 2A). The allowable molecule orientations were set to 0, 30, 60, …, and 330 degrees, which provides 12 orientations in total (molecule orientation is defined in the insert of Figure 2A). For reasons that will become clear later, the symmetry of the molecule is not used to reduce the number of possible orientations.

*Step 2. Molecule-surface interaction parameters.* For each adsorption site, an atomistic model consisting of an Au(111) slab and an FeFPc or MnHPc molecule positioned appropriately is generated. The molecule-surface interaction energy parameters $u_i$ are then obtained by performing single-point DFT calculations for each orientation (see Supporting Information 1 for a summary of the results). In practice, this step is only performed the adsorption sites located within a single unit cell, with the parameters for the other sites assigned according to symmetry.

*Step 3. Molecule-molecule interaction parameters.* Given the high density of adsorption sites per unit cell, multiple molecule orientations, and the two molecule types, the number of unique pairwise interactions that could occur on the surface is immense. Not only is it impractical to compute the energy for every case *via* DFT, it would also be inconvenient to search through the output of these calculations whenever $\varepsilon(c)$ is to be estimated *via* equation (1). We therefore assign the molecule-molecule interaction parameters on the basis of a machine-learned protocol (Fig 2B). This protocol accepts an atomistic model for a pairwise interaction as input and outputs a value for the interaction energy. The input atomistic model consists of a pair of molecules positioned according to two adsorption sites and orientations. In accordance with the assumptions of the Ising-like model, the atoms of the Au(111) slab are not included in this input. Upon input into the protocol, a set of features is extracted from the atomistic model, and those features are then passed through three machine-learned models in succession. The first model (SVM1) is a support vector machine which predicts whether the pairwise interaction is negligibly small ($|v_{ij}| < 0.05$ eV) or not. If the interaction is predicted to be negligible, then the protocol terminates and 0 eV is assigned as the interaction energy. Otherwise, the pairwise interaction is passed to a second support vector machine (SVM2), which predicts whether the interaction is 'repulsive' ($v_{ij} > 0$ eV) or 'attractive' ($v_{ij} < 0$ eV). If the interaction is predicted to be repulsive the protocol terminates and 10 eV is assigned as the interaction energy. If the interaction is predicted to be attractive, it is passed to a Gaussian process regression model (GPR), which assigns the final value of $v_{ij}$.

In this protocol, the features are formed by computing the eigenvalue spectrum of a symmetrized Coulomb matrix and projecting it onto a small set principal components. For the case of symmetric molecules, SVM1 and SVM2 were trained using datasets of 6300 and 3292 random pairwise interactions, respectively, and achieved fail rates of 1.5 % and 0.6 % respectively when compared to test data. GPR was trained using 3936 random pairwise interactions and demonstrated favorable predictive performance when compared to test data (see Figure 2C). Details on model training can be found in the Methods section and Supporting Information 2. The protocol could predict interaction energies within negligible computational time, making it suitable for assigning interaction energy parameters on the fly.

**3. Structure prediction with EUF+MCMC**

The advantage of the Ising-like model is that the Markov chain Monte Carlo (MCMC) method can be used to characterize the system at thermodynamic equilibrium. MCMC involves performing a random walk on the configuration space of the model in such a way that, over a long period of time, the number of visits by the random walk to configuration $c$ is proportional to the formation probability $p(c)$ in equation (2) [24]. In the usual conception of MCMC, one jump of the random walk corresponds to one of the molecules shifting to a neighboring adsorption site or changing its orientation. However, even though MCMC can be easily applied to the current model, in practice there is a major technical difficulty: the configuration space is both infinitely large (due to the infinitely large surface) and high dimensional (due to the two spatial dimensions per molecule). This combination is known to seriously reduce the convergence rate of Monte Carlo methods [24]; the random walk may need to make a considerable number of steps in order to reach the 'high probability region' (where all configurations have high formation probabilities), and may not even reach it within a practical computational time.

In order to address this problem, we employ Evolution Under Fire (EUF), a new type of genetic algorithm. EUF can be conceptualized as multiple random walks on the configuration space (Figure 3A). For a given random walk, each step corresponds to the exchange of individual molecules between isolated groups of molecules on the surface. EUF therefore ignores the long process of molecular diffusion between groups of molecules, and the random walks make much larger jumps across the configuration space compared the one used in MCMC. The large jump sizes, as well as a bias towards energy minimization, mean that the random walks quickly find the important configurations in which all molecules are tightly aggregated into a single group. Once one of these configurations is found, MCMC can be used to sample the configurations in its vicinity and characterize the molecular assembly at thermodynamic equilibrium (red arrows in Figure 3A).

Before introducing EUF, it is useful to recall how genetic algorithms are implemented under general settings unrelated to molecular self-assembly [25]. Consider $n$ variables $x_1, x_2, \ldots, x_n$, and an objective function $U(x_1, x_2, \ldots, x_n)$. The purpose of a genetic algorithm is to find the values of these variables for which $U$ is minimized. In step $i$ of a genetic algorithm, $m$ random vectors of the form

$v_1 = (x_{11}, x_{12}, \ldots, x_{1n})$,

$v_2 = (x_{21}, x_{22}, \ldots, x_{2n})$, ...

and

$v_m = (x_{m1}, x_{m2}, \ldots, x_{mn})$

are generated. The elements $x_{ij}$ are called genes, the random vectors are called chromosomes, and the $m$ chromosomes are collectively referred to as the population. In step $ii$, the fitness of each random vector is calculated. In this work, fitness is defined as

$$H_k = \frac{1 + U_{\max} - U_k}{1 + U_{\max} - U_{\min}} \tag{3}$$

where $k$ indicates chromosome $k$, $U_k$ is the value of the objective function computed using the genes in chromosome $k$, and $U_{\min}$ and $U_{\max}$ are the minimum and maximum objective function values in the population, respectively. In step $iii$ the $K$ new chromosomes are obtained by selecting pairs of chromosomes at random and conducting a so-called mating operation. Mating involves creating a new chromosome by randomly drawing genes from the two selected chromosomes. For example, if genes $v_1$ and $v_2$ are selected, then the new chromosome

$$v_{12} = (x_{11}, x_{22}, ..., x_{2n})$$

might result from their mating. Chromosomes are selected for breeding with a probability proportional to their fitness in (3). In step *iv*, the $N - K$ chromosomes with highest fitness in the original population are selected, and a new population consisting of these chromosomes and the *K* new ones from step *iii* is created. In step *v*, chromosomes are selected at random and random noise is added to their genes. This operation is called mutation. Steps *i* - *v* are iterated multiple times until $U_{min}$ reaches a sufficiently small value. Each chromosome can be regarded as the trajectory of a random walk with makes a transition at each iteration of steps *i – v*.

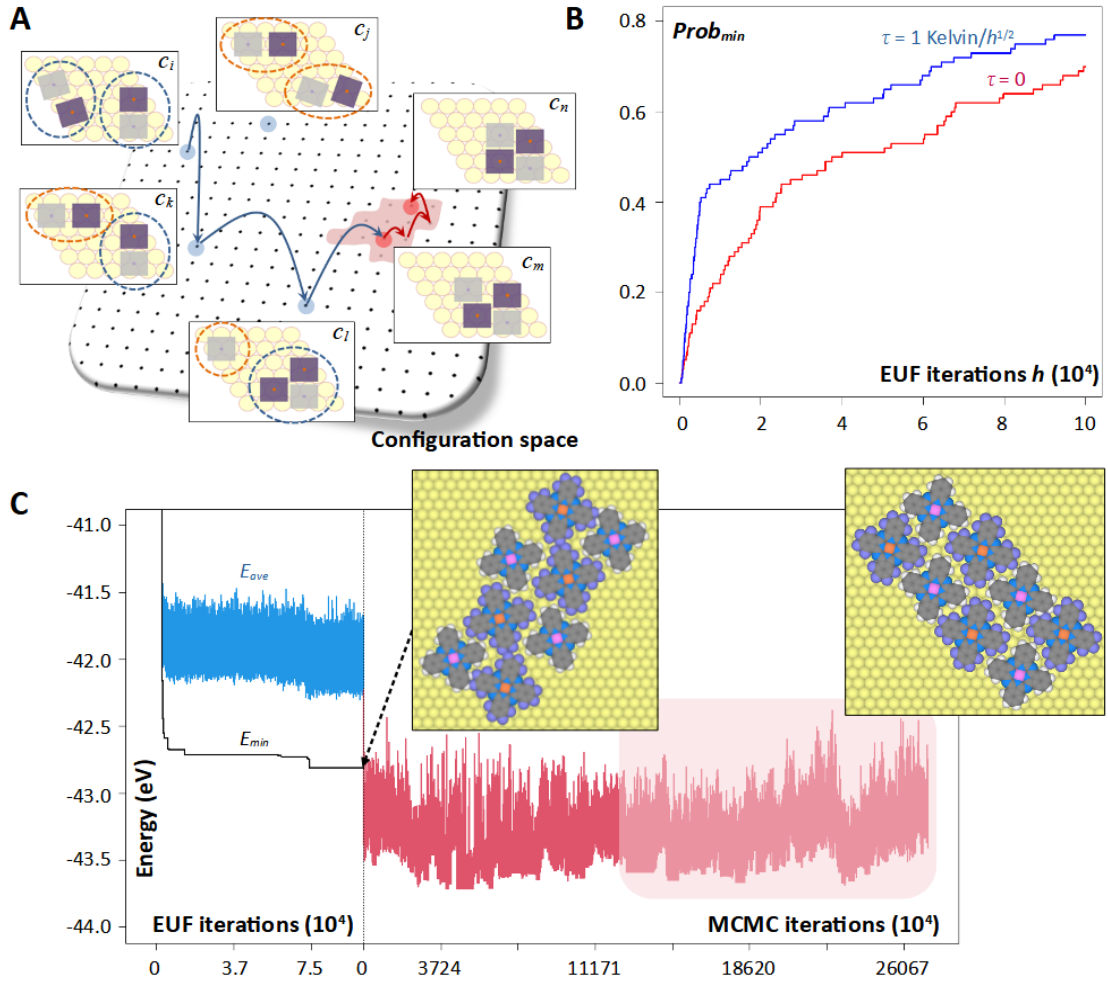

**Figure 3.** Summary of the EUF-MCMC method. (A) In essence, EUF ('evolution under fire') involves running multiple random walks on the configuration space of the model in Figure 2. In (A), the black points represent configurations in an abstract configuration space, and the blue arrows indicate the trajectory of one of these random walks. Each jump corresponds to mutation and breeding operations (see text for details). When the random walk reaches the 'high probability region' (red region), one can switch from EUF to Markov chain Monte Carlo (MCMC) to characterize the system at thermodynamic equilibrium. (B) The effect of annealing on the performance of EUF when applied to a system of adatoms. $Prob_{min}$ is the probability that the energy-minimizing adatom configuration is found. The red curve and blue curves represent the cases of no annealing and annealing. $\tau$ is the pseudotemperature (see text for details). (C) An example run of EUF+MCMC for the case of symmetric molecules on an Au(111) surface. The EUF phase is shown on left-hand side of the vertical dotted line and the MCMC phase is shown on the right-hand side. The EUF phase has been expanded for clarity. The blue line is the configuration energy averaged over all random walks, and the black line corresponds to the minimum energy configuration discovered so far. The subsequent MCMC run starts from this configuration (see middle insert). A typical configuration discovered towards the end of the MCMC run is shown in the right-hand insert.

In EUF, this scheme is adapted to describe the configurations of the statistical mechanical model. Rather than being vectors of numbers, the chromosomes consist of a vector of molecule subsets. Some examples of molecule subsets are shown in the inserts of Figure 3A (see the dotted lines). Each molecule subset corresponds to a group of molecules within which each molecule is undergoing a non-zero pairwise interaction with at least one other molecule from the group. Different subsets correspond to groups of widely separated molecules between which no non-zero interaction exists. For the objective function we choose a free energy expression

$$F(c) = \epsilon(c) - \tau S(c) \tag{4}$$

where *ε* is the energy computed with equation (1), *τ* is a positive parameter called the pseudotemperature, and *S* is the configuration entropy, defined as

$$S(c) = k_B \ln W(c) \tag{5}$$

where *W* is the weight of the configuration (the number of ways which the molecule subsets of the configuration can be placed onto the surface). An approximate formula for *W* is presented in Supporting Information 3. Mating operations are performed in an analogous way to that described above, with new chromosomes formed by drawing molecule subsets from the parent chromosomes at random. However, these operations are conditional on the numbers of molecules ($n_1$ and $n_2$) being conserved for each chromosome. The mutation operations in Step (v) are performed by selecting a molecule at random and changing its position and orientation at random, or shifting it to another molecule subset within the chromosome. In Figure 3A, the first transition (corresponding to the downward pointing blue arrow on the left-hand side) is the result of a mating operation between configurations $c_i$ and $c_j$. The second transition (from $c_k$ and $c_l$) is the result of a mutation operation.

EUF is named after the fact that it incorporates annealing to achieve a performance boost. This is illustrated in Figure 3B for a system consisting of 6 Na and 6 K adatoms on a Cu(111) surface (see Methods for implementation details). For this simple system, it is straightforward to find the energy-minimising adatom configuration with EUF alone. The red line in Figure 3B plots the probability of convergence to the global minimum, as computed from 100 independent runs of EUF with pseudotemperature *τ* set to zero. It can be seen that the probability of convergence quickly increases from 0 to about 0.5 within the first 40,000 iterations, after which it increases more slowly to a value of about 0.70 over the remaining iterations. The red curve in Figure 3B shows the case where $\tau = 1$ Kelvin/√(1 + *h*), where *h* is the number of iterations of EUF. In this work, *τ* is merely treated an algorithm parameter for tuning the performance of EUF, and does not have any physical significance. The strategy of gradually shrinking *τ* to zero with successive iterations is known as annealing, and serves to lift the chromosomes out of local energy minima. It can be seen that in the presence of annealing, the performance of the EUF is notably improved. In particular, the probability of convergence increases much more rapidly during the early part of the algorithm, reaching 0.5 after only 20,000 steps. After that it increases more slowly, reaching a value of about 0.8 by the end of the run. In the following, annealing is incorporated into all simulations into the EUF without further mention.

Figure 3C shows a complete application of the predictive method to a system of 4 FeFPc and 4MnHPc molecules adsorbed to an Au(111) surface at a temperature of 300 K. EUF was used for the first 100,000 iterations, and the minimum-energy configuration found during the EUF run was used as the initial condition of the subsequent MCMC run. It can be seen that by the end of its run, EUF found a molecular assembly consisting of two rows of alternating FeFPc and MnHPc molecules (middle insert). Over the subsequent long MCMC run, this prediction was refined and an assembly with high degree of order was identified as being important at thermodynamic equilibrium (right-hand side insert). In the following, this prediction method will be abbreviated as EUF+MCMC. In Supporting Information 4, these results are compared to a purely MCMC-based prediction which started from a random initial condition. During these MCMC simulations, a few of the molecules drifted far away from other others, and the system failed to converge to a sensible thermodynamic equilibrium state.

## 4. Self-assembly predictions

Ising-like were built for the cases of Au(111)-adsorbed symmetric (FeFPc and MnPc) and asymmetric (AFePc and AMnPc) molecules, with predictions carried out using EUF+MCMC. Each simulation used eight molecules in total. The t-distributed stochastic neighbor embedding (tSNE) method was used to visualize the MCMC sample [26]. Here, 50 configurations were selected at random from the second half of the MCMC trajectory, and for each pair of configurations a similarity metric was computed. This similarity metric was based upon the distances between metal ions in the molecular assemblies (see Methods). The result of tSNE is a two-dimensional scatter plot in which each point corresponds to one configuration, and clusters of points are similar according to the similarity metric. For practical purposes, the tSNE plot can be interpreted as a projection of the Boltzmann distribution in equation (2) onto a two-dimensional plane.

The result for a surface temperature of 100 K is shown in Figure 4A. Blue and maroon points correspond to the cases of symmetric and asymmetric molecules, respectively. For the case of symmetric molecules, the Boltzmann distribution is concentrated at two clusters, represented by the configurations $s_1$ and $s_2$, respectively. The configuration $s_1$ consists of tightly packed assemblies of alternating FeFPc and MnHPc molecules, strongly resembling the monolayer structure observed by Girovsky et al [22]. The distances between metal ions of neighboring molecules in $s_1$ range from 14.14 ± 0.07 Å, which compares well to the distances of 14.05 ± 0.08 Å observed in STM experiments [22]. A detailed comparison between configuration $s_1$ and experimental results is shown in Supporting Information 5. This agreement is impressive considering the grid and angle restrictions of the Ising-like model as well as the fact that the molecules were not allowed to relax. The other cluster corresponds to a more loosely packed phase represented by configuration $s_2$, for which the distances between metal ions of neighboring molecules ranges from 14.57 ± 0.16 Å.

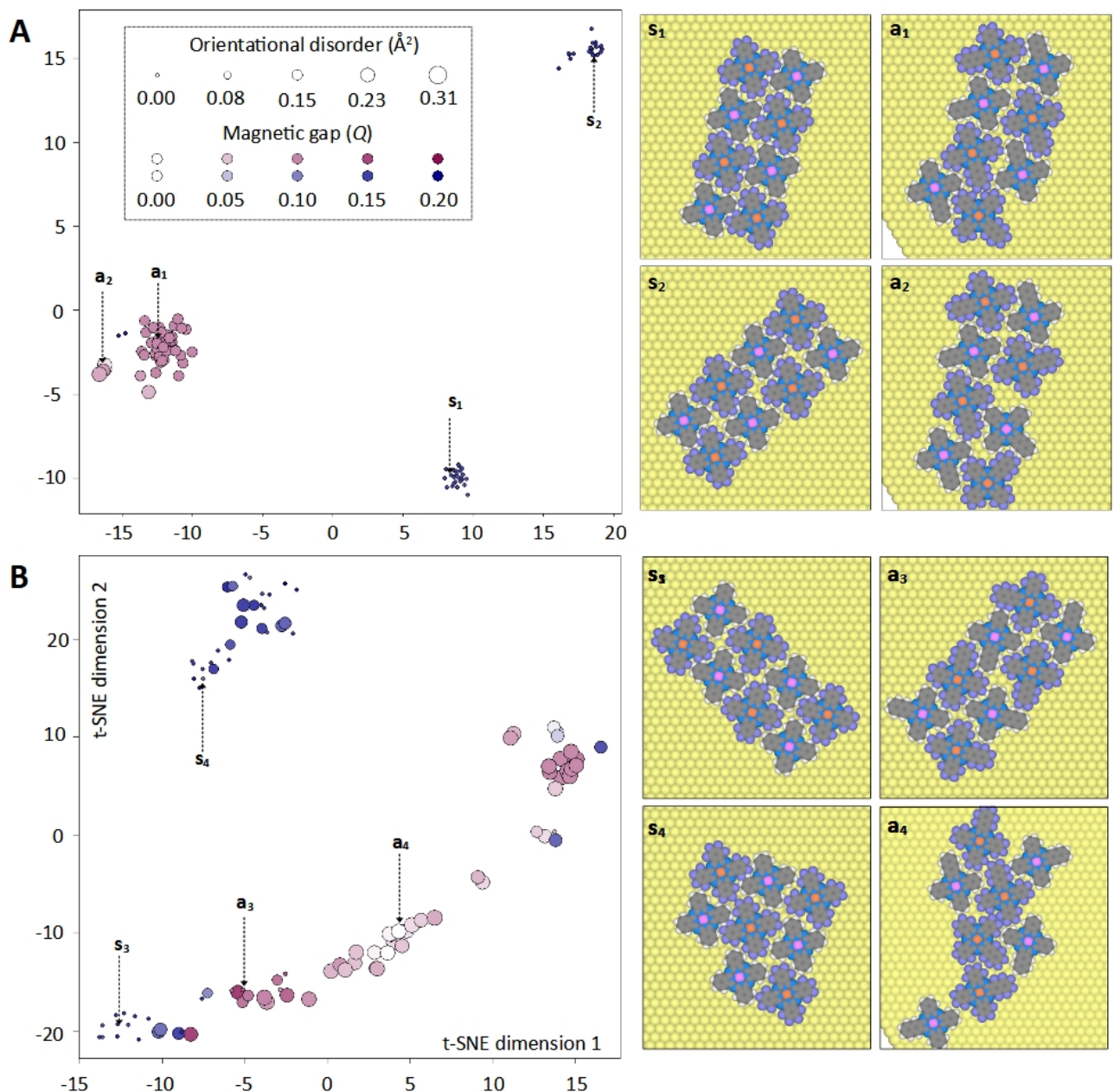

**Figure 4.** Characterization of the system at thermodynamic equilibrium using the t-distributed stochastic neighbor embedding (tSNE) method at (A) 100 K and (B) 300 K. Each point corresponds to one configuration, a few of which are shown on the right-hand side. Dark blue and maroon points are configurations for the case of symmetric and asymmetric molecules, respectively. The points are positioned so that nearby points have a similar arrangement of Fe and Mn ions. The radii of the points are proportional to an orientational disorder parameter, and the tone of the points corresponds to the magnetic gap of the molecular assembly (see text for details).

The situation is quite different for the case of the asymmetric molecules, in which the distribution is concentrated at only one cluster, as represented in Figure 4A by the configurations $a_1$ and $a_2$. The assemblies in $a_1$ and $a_2$ are similar to those in $s_1$ and $s_2$ in that they consist of two rows of alternating AFeFPc and AMnHPc molecules. However, in the asymmetric case, the degree of disorder in the assemblies appears to be higher, with the molecule orientations appearing to be less regular. Compared to the symmetric case, the AFeFPc and AMnHPc molecules appear to be in a state of agitation, constantly reorienting themselves but never settling into a specific, ordered arrangement.

The radii of the points in Figure 4A measure the degree of disorder in the orientations of the molecules in the configuration. Here, this 'orientational' disorder is calculated by taking each pair of molecules in the configuration, translating them so that their centers (metal ions) are coincident, and measuring the mean-squared distance between nearest atoms in the pair (see Methods for details). An orientational disorder of zero means that all pairs of molecules in the configuration are perfectly superimposable *via* translations alone. In the clusters to which $s_1$ and $s_2$ belong, the points are very small, indicating very little orientational disorder. The points in the cluster to which $a_1$ and $a_2$ belong are larger, indicating larger amounts of orientational disorder. However, for each of the three clusters observed in Figure 4A, the spatial spread of points is not large. Thus, while the assemblies formed from the asymmetric molecules display orientational disorder, the overall shapes of the assemblies are are relatively uniform within each cluster.

Figure 4B summarises the results obtained at room temperature (300 K). Again, the equilibrium distribution for the case of symmetric molecules is mainly concentrated on two clusters, which are represented by the configurations $s_3$ and $s_4$. While some of the configurations in these clusters exhibit orientational disorder, most points have very small radii, indicating that orientational order is maintained at room temperature. Moreover, while the two clusters show a higher spatial spread than the ones seen at 100 K, they remain relatively localised in space. This implies that the overall shapes of these assemblies are robust at room temperature. An important difference from the 100 K situation is that a cluster representing tightly packed assemblies ($s_1$) is not present. In particular, the cluster represented by $s_3$ appears to be very similar to the one represented by $s_2$ at 100 K. The loss of the tightly packed phase at higher temperatures is expected on entropic grounds, as tightly packed molecules do not any permit orientational disorder. The second cluster at 300 K (represented by $s_4$) is also a new feature compared to the 100 K case, and consists of molecules packed into a nearly square shaped assembly. These kinds of assembly have a vacancy at one corner, which arises from the use of an even number of molecules (eight) in these simulations. The vacancy represents an energetic instability, which prevents these assemblies from being observed at low temperatures.

The situation at room temperature is more interesting for the case of the asymmetric phthalocyanine molecules (Figure 4B, red points). Here, distinct clusters in the tSNE plot cannot be seen; rather, the points form a non-localised streak near the bottom-right of the image. At the left-hand side of this streak (represented by configuration $a_3$), a few points of low or moderate orientational disorder can be seen. These assemblies are similar to the ones seen at 100 K, in which the molecules are packed into two rows of alternating AFeFPc and AMnHPc molecules. However, as we move away from the left-hand edge of this streak configurations with very large orientational disorder become dominant. The assembly in configuration $a_4$, which was selected from the middle of this streak, is a representative example of these disordered configurations. At room temperature, the equilibrium dynamics of the asymmetric phthalocyanine molecular assemblies can therefore be described as *liquid-crystal-like*; the molecules constantly adjust their positions and orientations, and the assembly's overall shape is fluid, but occasionally they exhibit ordered configurations such as the one in $a_3$.

In order to explain the origin of this liquid crystal-like equilibrium dynamics, the surface-molecule and intermolecular interaction potentials for the symmetric and asymmetric cases were examined. For all four molecules, it was found that corrugation of the surface-molecule potential as a function of adsorption site (with fixed molecule orientation) was around 0.05 – 0.1 meV smaller per atom for the asymmetric case compared to the symmetric case (see Supporting Information 1). Moreover, for some adsorption sites, the corrugation of the surface-molecule interaction potential as a function of molecule orientation was similarly smaller for the asymmetric cases. The asymmetric ligand therefore has a slight smoothing effect on the surface-molecule potential, which may permit more disorder in the positions and orientations of the molecules in the assembly. However, it is difficult to attribute the liquid-crystal-like equilibrium dynamics to this small effect alone.

To examine whether there are any differences in the intermolecular interaction potentials of the symmetric and asymmetric cases, the distributions of the pairwise interaction energies were examined. These distributions were obtained from the training data used by the machine learning protocol in Figure 2B. For the case of symmetric molecules, the low-energy tail of the distribution was found to be sparsely populated; for interaction energies less than -0.30 eV, only two distinct pairwise interactions were observed (see Supporting Information 6). On the other hand, for the case of asymmetric molecules, the low energy tail of the distribution was found to be more densely populated; for interaction energies less than -0.30 eV, nine distinct pairwise interactions were observed. The origin of the liquid-crystal-like dynamics is therefore assigned to the availability of a larger number of low-energy pairwise interactions available in the asymmetric case, as well as to a small smoothing effect on the surface-molecule potential.

## 5. Magnetic ordering and potential spintronics applications

We now inquire into whether the liquid crystal-like equilibrium dynamics can act as a source of novel magnetic ordering in the asymmetric molecule assemblies. As was mentioned in the introduction, the magnetic ordering of Au(111)-adsorbed metal phthalocyanine monolayers arises from a competition between two effects: the RKKY effect, in which the coupling of metal ion spins of neighboring molecules is mediated by the surface electrons, and the Kondo effect, which screens the RKKY interaction [22, 23]. Interestingly, [23] reports that the Kondo effect for a metal ion spin becomes less pronounced as the number of neighboring molecules increases. For the molecular assemblies studied here, we would therefore expect for RKKY interactions to be stronger when both molecules in the pair have close neighbors located on all sides, and weaker then one of the molecules is near the edge of the assembly and has few immediate neighbors.

For any given molecular assembly, rigorous, first-principles predictions of the magnetic states of the metal ions are prohibitively costly. In order to gain some qualitative insights into the nature of the magnetic states, we instead employ a simplified Heisenberg-type model. For a given molecule configuration, the model Hamiltonian is given by

$$H = \sum_{i \neq j} J_{ij} \left( S_i^x S_j^x + S_i^y S_j^y + S_i^z S_j^z \right), \qquad (6)$$

where the sum is over all pairs of metal ion spins, $S_k^{\#}$, $\# = x, y$, or $z$, denotes the spin operator for the relevant component of metal ion spin $k$, and the $J_{ij}$ denotes a coupling coefficient. To incorporate RKKY interactions and Kondo screening, we define the coupling coefficients as

$$J_{ij} = Q\sigma_{ij}g\left(2k_F r_{ij}\right), \tag{7}$$

where $Q$ is a constant which depends upon the microscopic details of the electron system, $\sigma_{ij}$ is a screening factor, $k_F$ is the Fermi wavevector for Au(111), and $r_{ij}$ is the distance between metal ions. The function $g$ is defined according to the textbook formula for the distance dependence of the RKKY interaction [27]:

$$g(x) = \left(\frac{2}{x}\right)^4 (\sin x - x \cos x). \tag{8}$$

Because the literature does not provide any clear definition the screening factor $\sigma_{ij}$, we adopt the simple definition

$$\sigma_{ij} = \frac{1}{2}\left(\frac{N_i^{nn}}{4} + \frac{N_j^{nn}}{4}\right), \tag{9}$$

where $N^{nn}_k$ is the number of metal ions within a cut-off distance of metal ion $k$. Thus, screening will be absent ($\sigma_{ij} = 1$) when both molecules have neighbors on each of their four sides ($N^{nn}_k = 4$), and is effective ($0 \le \sigma_{ij} < 1$) otherwise. While the Hamiltonian in (6) makes assumptions which may not be justified for the current system (including isotropic coupling and the neglect of terms such as anisotropy [35]), it captures the essential physics which lead to magnetic ordering of Au(111)-adsorbed metal phthalocyanine monolayers, and is therefore appropriate for exploring how ligand asymmetry affects magnetic ordering.

In reference [22], antiferromagnetic ordering was observed in a monolayer of FeFPc and MnHPc after magnetization with a field oriented perpendicular to the surface. To study this situation, we suppose that one type of metal ion (either Fe or Mn) has a significantly slower relaxation time than the other, remaining aligned in the direction of the magnetic field after the latter is switch off. The spins of the other metal ions have a comparatively fast relaxation time and are free to relax. Let $T_1$ and $T_2$ denote the metal ions with fast and slow relaxation times, respectively. Under this situation, the Hamiltonian in (6) simplifies to

$$H_{\text{eff}} = \sum_{i \in T_1} \widetilde{h}_i S_i^z + \sum_{i,j \in T_1} J_{ij}\left(S_i^x S_j^x + S_i^y S_j^y + S_i^z S_j^z\right), \tag{10}$$

where the subscript 'eff' means 'effective', and

$$\widetilde{h}_i = s_2 \sum_{j \in T_2} J_{ij} \qquad (11)$$

denotes an effective magnetic field acting on spin $i$, which arises from the interaction with the metal ions of type $T_2$. In (11), $s_2$ is the magnitude of the spin for the $T_2$ metal ions. In writing (10) the constant additive term arising from interactions between $T_2$ ions has been ignored. Without loss of generality, the $T_1$ and $T_2$ are taken to be the Fe ions and Mn ions, respectively. The spin magnetic moments of Fe and Mn are taken to be 1 and 2, respectively, based on the values calculated in reference [22]. The cut-off distance for calculating (9) was set to 16 Å. $k_F$ is set to 0.18 Å$^{-1}$, which is the average of the two experimental values reported in [28]. $Q$ is left as an unspecified positive constant and the energies of the eigenstates ('magnetic states') computed from (10) are reported in units of $Q$.

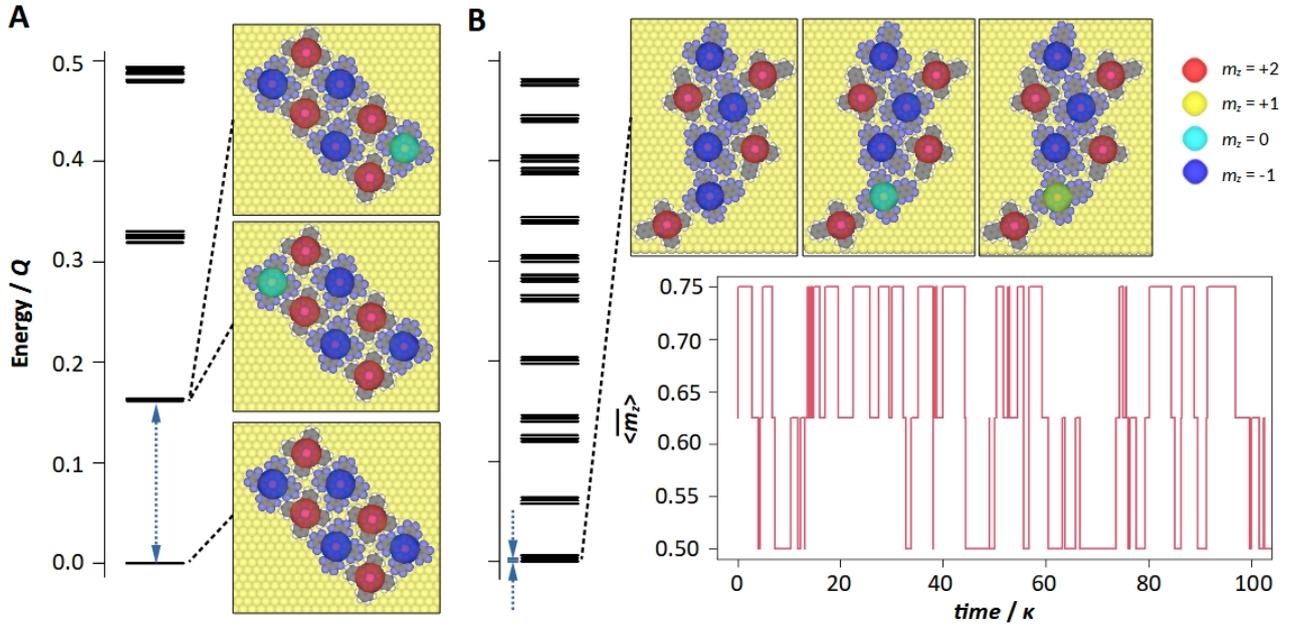

**Figure 5.** Low-energy magnetic states for the molecule configurations (A) $s_3$ and (B) $a_4$, as computed using a simple Heisenberg-type model with screened Ruderman-Kittle-Kasuya-Yoshida couplings between metal ion spins. Energy is reported in units of $Q$, the Ruderman-Kittel-Kasuya-Yoshida coupling parameter. In (A), the spin configurations which make the dominant contribution to the three lowest energy magnetic states are shown in the inserts. The $z$-components of the metal ion of the spins for each state are indicated by the coloured domes. The dotted blue arrows indicate the magnetic gap. In (B), the dominant spin configurations for the three lowest energy magnetic states are shown by the inserts at the top, in order of lowest energy (left) to highest energy (right). The plot at the bottom is a realisation of non-Gaussian (telegraph) noise simulated at low temperatures for configuration $a_4$. See text for further details.

In Figure 4A and 4B, the degree of shading of each point corresponds to the magnetic gap of the configuration. The magnetic gap is defined as the energy difference between the ground and first-excited magnetic states computed from (10). The lowest energy magnetic states for the configuration $s_3$ are shown in Figure 5A. The ground magnetic state shows a clear antiferromagnetic ordering, and the magnetic gap (blue arrow) is large. Thus, if this assembly was quenched to low temperatures we would expect for the ground spin state to be exclusively occupied and for the assembly to show antiferromagnetic ordering. Antiferromagnetic ordering and large magnetic gaps

(> 0.13 $Q$) are also exhibited by the other configurations for the symmetric molecules. The prediction of antiferromagnetic ordering for assemblies of symmetric phthalocyanine molecules is consistent with expectations based on experimental measurements [22].

We now turn to the spin ordering of the assemblies formed from the asymmetric phthalocyanine molecules. It can be seen from Figure 4B that the points for the asymmetric molecule cases are of a notably lighter shade compared to those of Figure 4A. In particular, several points in Figure 4B are essentially white, indicating a nearly zero magnetic gap. The low-energy magnetic states for one of these points (corresponding to the disordered configuration $a_4$) is shown in Figure 5B. It can be seen that at low energies, the magnetic states are arranged in tight groups of three. Within each group, the three states correspond to the flipping of the spin of one of the Fe ions located edge of the molecular assembly. The negligible energy difference between the three states of each group means that this spin flipping would easily occur as a result of ambient heat exchange with the surface. This spin flipping is due to the enhanced screening of RKKY interactions involving edge molecules, and is absent when screening is excluded from the model (see Supporting Information 7). However, this spin flipping however it is not a trivial consequence of this screening, because it was not observed for the assembly $s_3$. This spin flipping phenomenon is therefore the joint outcome of disorder in the assembly (driven by the liquid-crystal-like equilibrium dynamics) and screened RKKY interactions.

The presence of spin flipping suggests a novel application of these molecular assemblies: non-Gaussian noisy spin current modulators. It is well-known that spin currents can be transported through the surface states of Au(111) [29], and moreover these spin currents can be affected in various ways by the presence of magnetically ordered molecular assemblies (such as conversion from spin currents to charge currents [30]). At sufficiently low temperatures, the bottom three magnetic states of the assembly $a_4$ would be exclusively occupied. Ignoring the agitated flipping spin, the magnetic states display a stable magnetic order consisting of Fe ion spins oriented anti-parallel to the spins on the Mn ions. This magnetic ordering might be expected to modulate a spin current running through the surface states of the underlying Au(111) surface in some way. However it is modulated, telegraph noise – a quintessential example of non-Gaussian noise – would be superimposed on this modulation (see Figure 5B and the Methods section). This telegraph noise would arise from the spin flipping of the agitated electron.

The effects of Gaussian noise (as opposed to non-Gaussian noise) on nanodevice performance are thoroughly understood, having been characterised over a 60 year period spanning from Kubo's seminal work on the stochastic oscillator [31] through to on-going efforts to simulate open quantum systems [32]. However, Gaussian noise is an ideal situation, resulting from the collective outcome of a vast number of independent noise sources on the system. In the field of quantum information in particular, there is growing recognition that in many devices noise is decidedly non-Gaussian, which has stimulated an urgent need generate and characterise non-Gaussian noise sources on device performance [33]. The molecular assembly $a_4$ is therefore significant in this regard, as it could simultaneously act as a component of a spintronics device (due to its possession of magnetic order) while possessing a built-in source of non-Gaussian noise (due to the presence of the agitated spin). Telegraph noise is particularly natural type of non-Gaussian noise as well; as a close relative of the continuous-time random walk (which is known to converge to Gaussian noise in the limit of infinitely frequent transitions [34, 35]), it represents a kind of 'pre-Gaussian' noise regime which is arguably most relevant to device physics.

This prediction becomes particularly relevant when we reconsider the results of section 4, which suggest a direct pathway towards creating assemblies such as $a_4$. By depositing asymmetric phthalocyanine molecules onto Au(111) at room temperature and low surface coverages, we should be able to generate an ensemble of liquid-crystal-like molecular assemblies; and by quenching the system to a low temperature, we should be able to freeze-out disordered assemblies such as the one in $a_4$, which contain molecules with few nearest neighbors and hence undergoing weakened RKKY interactions on account of the Kondo effect.

## 6. Discussion and conclusions

Two scientific advancements will enable molecular self-assembly to be used as a routine method for building device components. Namely, the development of computational methods which can efficiently predict the outcome of the self-assembly process, and the development of strategies to realize molecular assemblies with collective physical properties. While this work has made important headway on both fronts, numerous points remain to be addressed.

The computational method presented here is effective because it embeds the complicated, many-body system into an Ising-like statistical mechanical model. However, by embedding the system into this model, four simplifications are made: (i) the molecules and the surface as treated as frozen objects with static internal degrees of freedom, (ii) the molecules are forced to reside on a grid of adsorption sites and can only adopt orientations from a finite set, (iii) the effect of the surface on intermolecular interactions is ignored, and (iv) third- and higher-order intermolecular interactions are ignored. In practice, simplification (i) limits the methodology to rigid and flat molecules such as porphyrins. However, this simplification is tolerable in view of the long-term goal of enabling molecular self-assembly as a methodology, which requires a well-defined starting point. Simplification (ii) can be alleviated by shrinking the step size of the grid and orientation set. This simplification therefore does not represent an inherent limitation of the computation method itself, just as how the use of finite $k$-point grids in DFT calculations does not represent an inherent limitation of the DFT method. Indeed, the good agreement with STM measursments obtained for the $s_1$ assembly with the grids used here is encouraging. Simplifications (iii) and (iv) are arguably more serious. For certain systems, including Cu(111)-adsorbed benzene and others [36, 37], it is well known that surface electrons immediately beneath the molecule become displaced and form a repulsive 'electron cloud' around the molecule. In turn, these electron clouds have a screening effect on intermolecular interactions, which is ignored in the energy expression in equation (1) (although it could perhaps be included by adding a correction term). While calculations on Cu(111)-adsorbed bianthracene molecules suggest that this screening does not affect the energetically preferred molecular assembly [12, 38], it is not clear whether these results hold for other types of molecules or surfaces as well. This is clearly an important point to address in future research. Similar remarks hold for simplification (iv), for which little is known about higher-order interactions between molecules on surfaces.

The major advantage of embedding the system into a Ising-like model is that the equilibrium state can be probed using MCMC methods. With the exception of molecular dynamics simulations, which are not applicable to these systems, MCMC is the only (routine) computational method which is guaranteed to produce a sample of configurations from the Boltzmann distribution. In view of the goal of enabling molecular self-assembly as a methodology, MCMC is highly advantageous because it allows us to explore how temperature affects the resulting assembly. Yet in order for MCMC to run efficiently an initial configuration close to those which dominate at equilibrium must be selected. In previous work this problem was dealt with by introducing a type of MCMC known as equivalence class sampling [39], however this method was problematic due to the approximate way in which entropy was handled. In the present paper I have reworked the ideas behind equivalence class sampling into the EUF (Evolution Under Fire) method. By using EUF to set the initial condition of the random walk in MCMC, we can greatly improve the efficiency of the MCMC method without sacrificing its accuracy. Future work aiming to accelerate prediction times should focus on computational implementations of this method (such as implementation on graphical processing units and quantum annealers).

While this computational method is effective, it demands massive overheads which acutely need to be addressed. Before predictions can be attempted using EUF+MCMC, numerous DFT calculations must be performed to tabulate the surface-molecule interaction energy terms ($u_i$ in equation 1), as well as to collect the training data for building the predictive models with machine learning (see Figure 2B). These overheads are highly prohibitive, as they are encountered each time a new type of molecule or surface is to be examined. Future work should aim at building highly general machine-learned models for the surface-molecule and molecule-molecule interactions using diverse molecule and surface datasets. Such models would not need to be re-fit whenever a new system is to be studied, mitigating the need to perform DFT calculations anew and permitting immediate predictions using EUF+MCMC.

In order to make headway towards the second scientific advance (strategies for creating molecular assemblies with collective physical properties) this work has focused on the specific case of magnetic order obtained from porphyrin-like metal complex molecules. For these systems, judicious structural modifications to the organic ligand can affect the type of molecular assembly obtained, which in turn controls the energetic ordering of the metal ion spin configurations. While this strategy is limited to porphyrin-like molecules adsorbed to metal surfaces, the virtually infinite range of structural modifications possible for these molecules offers a plethora of opportunities to material chemists. Indeed, combined with unsupervised learning techniques such as those from references [13] and [18], it may be possible to infer specific structural modifications which lead to a desired magnetic ordering. While magnetic ordering is one of the most important types of collective physical properties for device applications, the restriction of this strategy to porphyrin-like metal complex systems may be unsatisfying. On the other hand, it is difficult to imagine a magnetic ordering appearing in an assembly of purely organic molecules, without the presence of metal ions. In future work, the self-assembly of other classes of metal complex molecules, such as single-molecule magnets [40], may prove to be fruitful avenues for deducing new strategies for realizing novel magnetic order.

The specific strategy demonstrated here used an asymmetric ligand, which suppresses order in the resulting molecular assembly and endows it with liquid-crystal-like equilibrium dynamics. This disorder results gives rise to novel magnetic ordering, making them of potential use in spintronics and other applications. Structural asymmetry is an emerging theme in the inorganic chemistry literature, and is beginning to be explored in several fields [41]. The result obtained here is relevant because the asymmetric ligand is chemically feasible and the liquid-crystal-like assembly appears to be accessible at room temperature. Given their potential technological significance, it is important that these predictions are verified experimentally.

**Methods**

*Density functional theory calculations*

Density functional theory (DFT) calculations were performed as implemented in the Fritz Haber Institute ab initio molecular simulations (FHI-aims) package [42]. All calculations used default 'light' basis sets, the PBE exchange-correlation functional [43], Tkatchenko-Scheffler van der Waals corrections [44], and the atomic ZORA scalar relativity approximation [42]. DFT calculations for single molecules and pairs of molecules were performed without boundary conditions. DFT calculations involving Au(111) surfaces used 3-layer slab models, periodic boundary conditions, simulation cells with edge lengths of approximately 26 Å x 26 Å x 70 Å, and 1 x 1 x 1 k-point grids. Custom R scripts [45] were used to manage all high-throughput DFT calculations and to process results.

*Machine learning*

[Base dataset] For the two cases of symmetric and asymmetric molecules, 9000 pairwise interactions were generated at random using a custom R script, and the interaction energy computed for each one using DFT as described above.

[Feature extraction] For both cases of symmetric and asymmetric molecule samples, features were extracted in the following three steps. In the first step, symmetrized Coulomb matrices [46] were formed for each pairwise interaction in the sample. For pairwise interaction $k$ in a the sample, this matrix is defined as $\mathbf{C}_k = [C_{ij}^k]_{n \times n}$, where $n$ is the number of atoms per molecule and

$$C_{ij}^k = \frac{1}{2}\left(\frac{q_{ij}}{r_{ij}} + \frac{q_{ji}}{r_{ji}}\right). \tag{12}$$

Here, we have labeled one of the molecules in the pair as '1' and the other as '2'. In the above equation, $q_{ij}$ is the product of the atomic numbers of atom $i$ in molecule 1 and atom $j$ in molecule 2, and $r_{ij}$ is the distance between atom $i$ of molecule 1 and atom $j$ of molecule 2. In the second step, for each pairwise interaction $k$, the eigenvalue spectrum of $\mathbf{C}_k$ was calculated. This resulted in a 9000 x $n$ data matrix, where 9000 is the size of the original dataset. In the final step, principal component analysis was performed on this data matrix. For each pairwise interaction, the projection of its eigenvalue spectrum onto the first six principal components was computed, and those six projections were used as features in the subsequent steps. Principal component analysis was performed using the prcomp function in R [45].

[Support vector machines] The support vector machines SVM1 and SVM2 in Figure 2 were built using the LIBSVM 3.24 library for C++ [47]. Both cases used slack variables with cost parameter $C$ and radial basis kernels with inverse length scale $\gamma$. Parameters $C$ and $\gamma$ were optimized by grid search, where the optimal parameters were those which minimized the average SVM fail rate from three-fold cross validation. For both cases, 70 % of the original data was used as training data. For the case of SVM2, pairwise interactions which had energies between -0.05 and 0.05 eV were removed from the original dataset before selecting the training data. For the case of symmetric molecules, SVM1 and SVM2 achieved average fail rates of 1.5 % and 0.6 %, respectively during cross validation. For the case of asymmetric molecules, the average fail rates were 2.1 % and 1.5 %, respectively.

[Gaussian process regression] Gaussian process regression models were built using custom C++ codes. For both cases of symmetric and asymmetric molecules, the covariance matrix $\mathbf{\Sigma} = \mathbf{K} + \sigma^2 \mathbf{I}$ was used. Here, $\mathbf{I}$ is the $m \times m$ identity matrix, where $m$ is the number of training data points, and $\mathbf{K} = [K_{ij}]_{m \times m}$, where $i$ and $j$ denote two pairwise interactions from the training data, $K_{ij} = a \exp(-\mathbf{b} \cdot (\mathbf{q}_i - \mathbf{q}_j)^2)$, and

$$\mathbf{b} \cdot (\mathbf{q}_i - \mathbf{q}_j)^2 = \sum_{k=1}^{d} b_k \left( q_k^i - q_k^j \right)^2, \tag{13}$$

where $d$ is the feature dimension ($d = 6$ in this work), and $q_k^i$ is the $k$th feature for pairwise interaction $i$ in the sample. The hyperparameters $a$, $b_1$, $b_2$,…, $b_d$, and $\sigma$ were optimised using five-fold cross validation on the training data, where the optimal parameters were the ones which minimised the average of the mean-square errors computed from each test fold. This optimisation was performed using the limited-memory Broyden-Fletcher-Goldfarb-Shanno algorithm [48] as implemented in R. The term $\sigma \mathbf{I}$ was included to improve numerical stability when inverting the covariance matrix. For the case of symmetric molecules, 3936 pairwise interactions from the original sample were used as training data points and 695 were used as test points. For the case asymmetric molecules, 3943 points were used as training data and 696 points were used as test data. In both cases, pairwise interactions with interaction energy greater than -0.05 eV were excluded when selecting the training and test points from the original data. The good performance of the resulting GPR models against the test data is shown in Supporting Information 2.

*Evolution Under Fire*

The Evolution Under Fire (EUF) algorithm was implemented using a custom C++ code. All EUF runs used 100 chromosomes, pseudotemperature $\tau = 1.0$ Kelvin / $\sqrt{h}$, where $h$ is the iteration number, and configuration weight $W$ as described in Supporting Information 3. At each iteration, 50 chromosomes were selected to survive, and the remaining 50 chromosomes were generated by breeding operations. Chromosomes were selected for mutation with probability 0.05.

*Markov Chain Monte Carlo*

Markov chain Monte Carlo (MCMC) was performed with the standard Metropolis-Hastings algorithm with parallel tempering, and was implemented using a custom C++ code. MCMC 'moves' were performed by selecting a molecule at random and shifting it to a neighboring adsorption site at random and changing its orientation by -30º, 0º, or 30º. Parallel tempering was implemented with 10 replicates. For the 100 K simulations, the replicate temperatures ranged from 100 K to 500 K. For the 300 K simulations, replicate temperatures ranged from 300 K to 1000 K.

*EUF benchmarking for Cu(111)-adsorbed Na/K atoms*

DFT calculations were performed using the setting described above, however with the PW-LDA exchange-correlation functional [49] and without van der Waals corrections. Calculations involving Cu(111) slabs used 4-layer slabs, periodic boundary conditions, simulation cells with edge lengths of approximately 10 Å x 10 Å x 40 Å, and 2 x 2 x 1 k-point grids. DFT calculations for Na-Na, Na-K, and K-K interactions energies were performed without boundary conditions.

A dataset of 750 random interatomic interactions (Na-Na, Na-K, and K-K) was generated, 375 of which were used as training data to build a GPR model based upon interatomic distances and atomic numbers as features. Near perfect agreement with the remaining 375 training data points was obtained. Support vector machine models were not used when assigning interaction energies to configurations.

EUF was repeated 100 times, each time using the same setting as described above. The output was analysed with a custom R script. $Prob_{min}$ was defined as the fraction of the 100 EUF runs for which the minimum-energy chromosome had energy less than -45 eV. The adsorption site grid had 16 sites per Cu(111) unit cell. The parameter $N$ in the configuration weight (see Supporting Information 3) was set to 225.

*t-Distributed stochastic neighbor embedding (tSNE)*

tSNE was performed as implemented in the Rtsne package [50] for R [45], with the perplexity parameter set to 10. The jitter function (with the "amount" parameter set to 0.5) was used to reduce overlapping of points.

*Orientational disorder parameter*

For a given configuration, the orientational order parameter was calculated as follows. Consider two molecules $r$ and $s$ from $c$, and translate $r$ so that its metal ion coincides with that of $s$. Define

$$w_{rs} = \frac{1}{m} \sum_{i=1}^{m} d_{i,s}, \qquad (14)$$

where $m$ is the number of atoms in molecule $r$, and $d_{i,s}$ is the distance between atom $i$ in molecule $r$ and the nearest atom from $i$ in molecule $s$. Note that $w_{rs}$ does not necessarily equal $w_{sr}$. The orientational disorder for configuration $c$ is defined as

$$\Omega(c) = \frac{1}{n(n-1)} \sum_{r \neq s} (W_{rs} - \bar{W})^2 \tag{15}$$

where $n$ is the number of molecules in $c$, $W_{rs} = (w_{rs} + w_{sr})/2$ and $\bar{W}$ is the mean of $W_{rs}$ over all pairs of molecules.

*Simulation of telegraph noise*

Telegraph noise was simulated as a random walk on the eigenstates of the Hamiltonian in (10) using a custom R script. The transition frequency from eigenstate $i$ to eigenstate $j$ was set to $\kappa^{-1} P_i/P_j$, where $\kappa$ is a constant (in units of time), $P_k \propto \exp(-\varepsilon_k/T)$, $\varepsilon_k$ is the energy of eigenstate $k$, and $T$ is proportional to temperature. Transitions between eigenstates were only permitted if the $z$-component of the total magnetic moment $M_Z$ of the Fe ions (as computed from the expectation value of $S_1^z + S_2^z + S_3^z + S_4^z$) changed by exactly $\pm 1$ (this corresponds to the molecular assembly exchanging units of angular momentum with a heat bath). For the simulation shown in Figure 5B, $T$ was set to 0.01 (in units of $Q$, as defined in equation (7)). The initial state was also sampled according to the probability $P_k \propto \exp(-\varepsilon_k/T)$. Due to the small value of $T$, the random walk only explored the three lowest energy states during the simulation. In Figure 5B, the telegraph noise is reported as the $z$-component of the total magnetic moment of the entire system (including both Fe and Mn ions) averaged over the 8 ions (i.e., $\langle \bar{m}_z \rangle = (M_Z/4 + 8/4)/2$).

**Acknowledgements**

This research was supported by JSPS KAKENHI grant 19H04574 for the project area "Coordination Asymmetry: Design of Asymmetric Coordination Sphere and Anisotropic Assembly for the Creation of Functional Molecules" and JSPS KAKENHI grant 21K05003 for scientific research (C).

**References**

[1] Butler, S. *Darwin Among the Machines.* The Press, Christchurch, New Zealand, 1863.

[2] Whitesides, G. M., Boncheva, M. Beyond molecules: Self-assembly of mesoscopic and macroscopic components. *Proc. Natl. Acad. Sci. U.S.A.* **99**, 2002, 4773.

[3] Wakayama, Y. On-surface molecular nanoarchitechtonics: From self-assembly to directed assembly. *Jpn. J. Appl. Phys.* **55**, 2016, 1102AA

[4] Goronzy, D. P., Ebrahimi, M., Rosei, F., Arramel, Fang, Y., De Feyter, S., Tait, S. L., Wang, C., Beton, P. H., Wee, A. T. S., Weiss, P. S., Perepichka, D. F. Supramolecular Assemblies on Surfaces: Nanopatterning, Functionality, and Reactivity. *ACS Nano*, **12**, 2018, 7445.

[5] Ariga, K., Nishikawa, M., Mori, T., Takeya, J., Shrestha, L. K., and Hill, J. P. Self-assembly as a key player for materials nanoarchitechtonics. *Sci. Technol. Adv. Mater.* **20**, 2019, 52.

[6] Talapin, D. V., Engel, M., and Braun, P. V. Functional materials and devices by self-assembly. *MRS Bull.* **45**, 2020, 799

[7] Wang, J., Li, H., Xu, B. Biological functions of supramolecular assemblies of small molecules in the cellular environment. *RSC Chem. Biol.* **2**, 2021, 289

[8] Takezawa, H., Fujita, M. Molecular Confinement Effects by Self-Assembled Coordination Cages. *Bull. Chem. Soc. Jpn.* **94**, 2021, 2351.

[9] Tao, L., Zhang, Y-Y., Du, S. Structures and electronic properties of functional molecules on metal substrates: From single molecule to self-assemblies. *WIREs Comput. Mol. Sci.* 2021, e1591.

[10] Todorovic, M., Gutmann, M. U., Corander, J., Rinke, P. Bayesian inference of atomistic structure in functional materials. *npj Comput. Mater.* **5**, 2019, 35.

[11] Jarvi, J., Alldritt, B., Krejci, O., Todorovic, M., Liljeroth, P., Rinke, P. Integrating Bayesian Inference with Scanning Probe Experiments for Robust Identification of Surface Adsorbate Configurations. *Adv. Funct. Mater.* **31**, 2021, 2010853.

[12] Packwood, D. M., Han, P., Hitosugi, T. Chemical and entropic control on the molecular self-assembly process. *Nat. Commun.* **8**, 2018, 14463.

[13] Packwood, D. M., Hitosugi, T. Materials informatics for self-assembly of functionalized organic precursors on metal surfaces. *Nat. Commun.* **9**, 2018, 2469

[14] Packwood, D. M., Hitosugi, T. Rapid prediction of molecule arrangements on metal surfaces *via* Bayesian optimization. *Appl. Phys. Express.* **10**, 2017, 065502

[15] Hormann, L., Jeindl, A., Egger, A. T., Scherbela, M., Hofmann, O. T. SAMPLE: Surface structure search enabled by coarse graining and statistical learning. *Comput. Phys. Commun.* **244**, 2019, 143.

[16] Obersteiner, V., Scherbela, M., Hormann, L., Wegner, D., Hofmann, O. T. Structure Prediction for Surface-Induced Phases of Organic Monolayers: Overcoming the Computational Bottleneck. *Nano Lett.* **17**, 2017, 4453.

[17] Egger, A. T., Hormann, L., Jeindl, A., Scherbela, M., Obersteiner, V., Todorovic, M., Rinke, P., Hofmann, O. T. *Adv. Sci.* **7**, 2020, 2000992

[18] Packwood, D. M. Exploring the configuration spaces of surface materials using time-dependent diffraction patterns and unsupervised learning. *Sci. Rep.* **10**, 2020, 5868.

[19] Sk, R., Deshpande, A. Unveiling the emergence of functional materials with STM: metal phthalocyanine on surface architectures. *Mol. Syst. Des. Eng.* **4**, 2019, 471.

[20] Kumar, K. S., Ruben, M. Sublimable Spin-Crossover Complexes: From Spin-State Switching to Molecular Devices. *Angew. Chem. Int. Ed.* **60**, 2021, 7502.

[21] Meng, T., Lei, P., Zeng, Q. Progress in the self-assembly of porphyrin derivatives on surfaces: STM reveals. *New J. Chem.* **45**, 2021, 15739.

[22] Girovsky, J., Nowakowski, J., Ali, Md. E., Baljozovic, M., Rossmann, H. R., Nijs, T., Aeby, E. A., Nowakowska, S., Siewert, D., Srivastava, G., Wackerlin, C., Dreiser, J., Decurtins, S., Liu, S-X., Oppeneer, P. M., Jung, T. A., Ballav, N. Long-range ferrimagnetic order in a two-dimensional supramolecular Kondo lattice. *Nat. Commun.* **8**, 2017, 15388.

[23] Tsukahara, N., Shiraki, S., Itou, S., Ohta, N., Takagi, N., Kawai, M. Evolution of Kondo Resonance from a Single Impurity Molecule to the Two-Dimensional Lattice. *Phys. Rev. Lett.* 106, 2011, 187201.

[24] Robert, C. L., Casella, G. *Monte Carlo Statistical Methods.* Springer, New York, New York (2004)

[25] Zwillinger, D. *CRC Standard Mathematical Tables and Formula* (32$^{nd}$ Edition). CRC Press, Boca Radon, Florida (2012).

[26] van der Maaten, L. Accelerating t-SNE using a tree-based algorithm. *J. Mach. Learn. Res.* **15**, 2014, 1.

[27] Patterson, J. D., Bailey, B. C. *Solid-State Physics Introduction to the Theory.* Springer, New York, New York (2007)

[28] Reinert, F., Nicolay, G., Schmidt, S., Ehm, S. D., Hufner, S. Direct measurement of the L-gap surface states of the (111) face of nobel metals by photoelectron spectroscopy. *Phys. Rev. B.* **63**, 2001, 115415.

[29] Liu, M-H., Chen, S-H., Chang, C-R. Nonequilibrium spin transport on Au(111) surfaces. *Pys. Rev. B.* **78**, 2008, 195413.

[30] Isshiki,H., Kondou, K., Takizawa, S., Shimose, K., Kawabe, T., Minamitani, E., Yamaguchi, N., Ishii, F., Shiotari, A., Sugimoto, Y., Miwa, S., Otani, Y. Realization of Spin-Dependent Functionality by Covering a Metal Surface with a Single Layer of Molecules. *Nano. Lett.* **19**, 2019, 7119.

[31] Kubo, R. A Stochastic Theory of Line Shape. *Adv. Chem. Phys.* **15**, 1969, 101.

[32] Tanimura, Y. Stochastic Liouville, Langevin, Fokker-Planck, and Master Equation Approaches to Quantum Dissipative Systems. *J. Phys. Soc. Jpn.* **75**, 2006, 082001.

[33] Sung, Y., Beaudoin, F., Norris, L. M., Yan, F., Kim, D. K., Qiu, J. Y., von Lupke, U., Yoder, J. L., Orlando, T. P., Gustavsson, S., Viola, L., Oliver, W. D. Non-Gaussian noise spectroscopy with a superconducting qubit sensor. *Nat. Commun.* **10**, 2019, 3715.

[34] Packwood, D. M., Tanimura, Y. Non-Gaussian stochastic dynamics of spins and oscillators: A continuous-time random walk approach. *Phys. Rev. E.* **84**, 2011, 061111.

[35] Packwood, D. M., Tanimura, Y. Dephasing by a continuous-time random walk process. *Phys. Rev. E.* **86**, 2012, 011130.

[36] Bagus, P. S., Hermann, K., Woll, C. The interaction of $C_6H_6$ and $C_6H_{12}$ with noble metal surfaces: electronic level alignment and the origin of the interface dipole. *J. Chem. Phys.* **123**, 2005, 184109.

[37] Bagus, S., Staemmler, V., Woll, C. Exchange-like Effects for Closed-Shell Adsorbates: Interface Dipole and Work Function. *Phys. Rev. Lett.* **89**, 2002, 096104.

[38] Li, X., Packwood, D. M. Substrate-molecule decoupling induced by self-assembly – Implications for graphene nanoribbon fabrication. *AIP Adv.* **8**, 2018, 045117.

[39] Packwood, D. M., Han, P., Hitosugi, T. State-space reduction and equivalence class sampling for a molecular self-assembly model. *R. Soc. Open. Sci.* **3**, 2016, 150681.

[40] Gatteschi, D., Sessoli, R., Villian, J. *Molecular Nanomagnets*. Oxford University Press, Oxford, U. K. (2006)

[41] Ariga, K and Shionoya, M. Nanoarchitectonics for Coordination Asymmetry and Related Chemistry. *Bull. Chem. Soc. Jpn.* **94**, 2021, 839.

[42] Blum, V., Gehrke, R., Hanke, F., Havu, P., Havu, V., Ren, X., Reuter, K., Scheffler, M. *Ab initio molecular simulations with numeric atom-centered orbitals. Comput. Phys. Commun.* **180**, 2009, 2175.

[43] Perdew, J. P., Burke, K., Ernzerhof, M. Generalized Gradient Approximation Made Simple. *Phys. Rev. Lett.* **77**, 1996, 3865.

[44] Tkatchenko, A., Scheffler, M. Accurate Molecular Van Der Waals Interactions from Ground-State Electron Density and Free-Atom Reference Data. *Phys. Rev. Lett.* **102**, 2009, 073005.

[45] R Core Team. R: A language and environment for statistical computing. R Foundation for Statistical Computing, Vienna, Austria. https://www.r-project.org/

[46] Rupp, M., Tkatchenko, A., Muller, K-R., von Lilienfeld, A. Fast and Accurate Modeling of Molecular Atomization Energies with Machine Learning. *Phys. Rev. Lett.* **108**, 2012, 058301.

[47] Chang, C-C., Lin, C-J. LIBSVM: a library for support vectory machines. *ACM Trans. Intell. Syst. Technol.* **2**, 2011, 1. Software available at http://www.csie.ntu.edu.tw/~cjlin/libsvm

[48] Byrd, R. H., Lu, P., Nocedal, J., Zhu, C. A limited memory algorithm for bound constrained optimization. *SIAM J. Sci. Comput.* **16**, 1995, 1190.

[49] J. P. Perdew and Y. Wang. Accurate and simple analytic representation of the electron-gas correlation energy. *Phys. Rev. B*, **45**:13244–13249, 1992

[50] Krijthe, J. H. Rtsne: T-Distributed Stochastic Neighbor Embedding using a Barnes-Hut Implementation. R package version 0.15. https://github.com/jkrijthe/Rtsne

[51] Momma, K., Izumi, F. VESTA 3 for three-dimensional visualization of crystal, volumetric, and morphology data. *J. Appl. Crystallogr.* **44**, 2011, 1272.

**Supporting information for**

**Magnetic on-surface assemblies predicted from a pious computational method**

Daniel M. Packwood

Institute for Integrated Cell-Material Sciences (iCeMS), Kyoto University, Kyoto 606-8502, Japan

**Supporting information 1. Summary of surface-molecule adsorption energies**

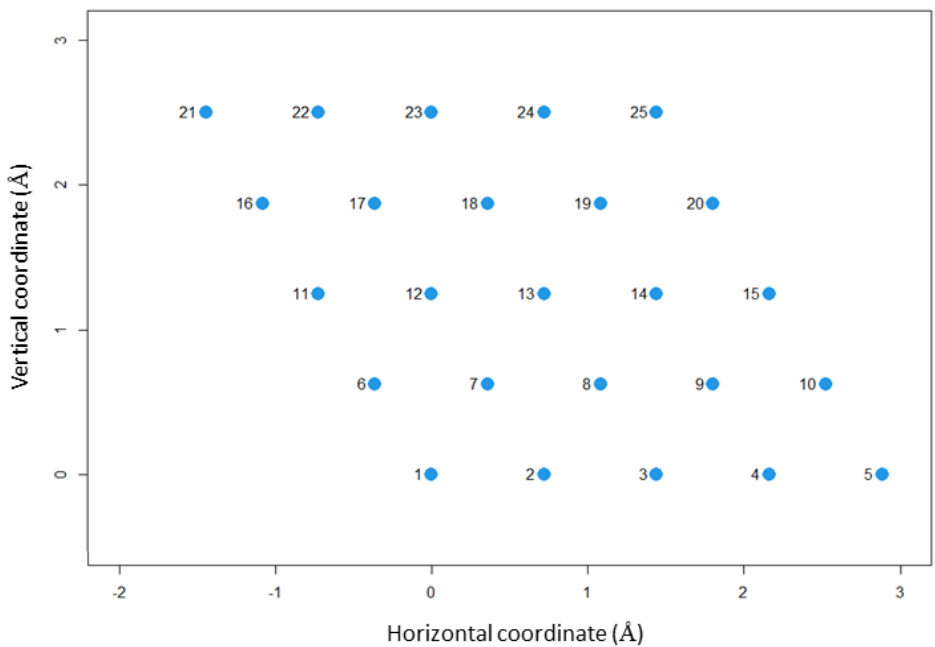

Plot of the grid points (adsorption sites) contained within a single Au(111) unit cell.

|        | 0º    | 30º   | 60º   | 90º   | 120º  | 150º  | 180º  | 210º  | 240º  | 270º  | 300º  | 330º  |
|--------|-------|-------|-------|-------|-------|-------|-------|-------|-------|-------|-------|-------|
| $m$    | -5.02 | -5.02 | -5.03 | -5.02 | -5.02 | -5.02 | -5.02 | -5.02 | -5.03 | -5.02 | -5.02 | -5.03 |
| $s$    | 0.035 | 0.035 | 0.035 | 0.035 | 0.035 | 0.035 | 0.035 | 0.035 | 0.035 | 0.035 | 0.035 | 0.035 |
| $m/n_{at}$ | -8.81 | -8.81 | -8.82 | -8.81 | -8.81 | -8.82 | -8.81 | -8.81 | -8.82 | -8.81 | -8.81 | -8.82 |
| $s/n_{at}$ | 0.062 | 0.062 | 0.061 | 0.062 | 0.062 | 0.061 | 0.062 | 0.062 | 0.061 | 0.062 | 0.062 | 0.061 |

Summary statistics for the adsorption energies of a FeFPc molecule in various orientations on Au(111). $m$ is the energy averaged over adsorption sites in units of eV. $s$ is the standard deviation in units of eV. $m$ and $s$ measure the strength and corrugation of the surface-molecule interaction potential for a fixed molecule orientation, respectively. $n_{at}$ is the number of atoms in the molecule, and $m/n_{at}$ and $s/n_{at}$ are per-atom average energy and standard deviation, respectively, in units of $10^{-2}$ eV.

|        | 0º    | 30º   | 60º   | 90º   | 120º  | 150º  | 180º  | 210º  | 240º  | 270º  | 300º  | 330º  |
|--------|-------|-------|-------|-------|-------|-------|-------|-------|-------|-------|-------|-------|
| $m$    | -5.45 | -5.45 | -5.45 | -5.45 | -5.45 | -5.45 | -5.45 | -5.45 | -5.45 | -5.45 | -5.45 | -5.45 |
| $s$    | 0.035 | 0.036 | 0.034 | 0.037 | 0.035 | 0.036 | 0.035 | 0.037 | 0.035 | 0.036 | 0.034 | 0.036 |
| $m/n_{at}$ | -8.65 | -8.65 | -8.65 | 8.65 | -8.65 | -8.65 | -8.65 | -8.65 | -8.65 | -8.65 | -8.65 | -8.65 |
| $s/n_{at}$ | 0.056 | 0.057 | 0.054 | 0.058 | 0.055 | 0.057 | 0.055 | 0.058 | 0.055 | 0.057 | 0.054 | 0.058 |

As above, but for an AFeFPc molecule. Shaded cells indicate orientations for which the corrugation of the surface-molecule interaction potential (as measured by $s/n_{at}$) decreased relative to the case of FeFPc.

|     | 0°   | 30°  | 60°  | 90°  | 120° | 150° | 180° | 210° | 240° | 270° | 300° | 330° |
|-----|------|------|------|------|------|------|------|------|------|------|------|------|
| $m$ | -5.05 | -5.05 | -5.05 | -5.05 | -5.05 | -5.05 | -5.05 | -5.05 | -5.05 | -5.05 | -5.05 | -5.05 |
| $s$ | 0.061 | 0.062 | 0.061 | 0.061 | 0.062 | 0.061 | 0.061 | 0.062 | 0.061 | 0.061 | 0.062 | 0.061 |
| $m/n_{at}$ | -8.87 | -8.85 | -8.87 | -8.87 | -8.85 | -8.87 | -8.87 | -8.85 | -8.87 | -8.87 | -8.85 | -8.87 |
| $s/n_{at}$ | 0.107 | 0.108 | 0.107 | 0.107 | 0.108 | 0.107 | 0.107 | 0.108 | 0.107 | 0.107 | 0.108 | 0.107 |

As above, but for a MnHPc molecule.

|     | 0°   | 30°  | 60°  | 90°  | 120° | 150° | 180° | 210° | 240° | 270° | 300° | 330° |
|-----|------|------|------|------|------|------|------|------|------|------|------|------|
| $m$ | -5.52 | -5.51 | -5.52 | -5.52 | -5.51 | -5.52 | -5.52 | -5.51 | -5.52 | -5.52 | -5.51 | -5.52 |
| $s$ | 0.060 | 0.060 | 0.057 | 0.061 | 0.060 | 0.060 | 0.058 | 0.062 | 0.058 | 0.059 | 0.058 | 0.062 |
| $m/n_{at}$ | -8.76 | -8.75 | -8.76 | -8.76 | -8.75 | -8.76 | -8.76 | -8.75 | -8.76 | -8.76 | -8.75 | -8.76 |
| $s/n_{at}$ | 0.095 | 0.096 | 0.090 | 0.096 | 0.095 | 0.095 | 0.092 | 0.098 | 0.092 | 0.094 | 0.093 | 0.098 |

As above, but for an AMnHPc molecule. Shaded cells indicate orientations for which the corrugation of the surface-molecule interaction potential (as measured by $s/n_{at}$) decreased relative to the case of MnHPc.

| Site | 1 | 3 | 6 | 8 | 11 | 13 | 16 | 18 |
|------|---|---|---|---|----|----|----|----|
| $m$ | -5.08 | -4.99 | -5.04 | -4.99 | -4.99 | -4.99 | -5.04 | -4.99 |
| $s$ | 0.00037 | 0.014 | 0.0065 | 0.0074 | 0.015 | 0.015 | 0.0064 | 0.0067 |
| $m/n_{at}$ | -8.92 | -8.76 | -8.84 | -8.76 | -8.75 | -8.75 | -8.84 | -8.76 |
| $s/n_{at}$ | 0.00065 | 0.025 | 0.011 | 0.013 | 0.026 | 0.026 | 0.011 | 0.012 |

Summary statistics for the adsorption energies of a FeFPc molecule at various adsorption sites on Au(111). $m$ is the energy averaged over all orientations in units of eV. $s$ is the standard deviation in units of eV. $m$ and $s$ measure the strength and corrugation of the surface-molecule interaction potential for a fixed adsorption site, respectively. $m/n_{at}$ and $s/n_{at}$ are par-atom average energy and standard deviation, respectively, in units of $10^{-2}$ eV.

| Site | 1 | 3 | 6 | 8 | 11 | 13 | 16 | 18 |
|------|---|---|---|---|----|----|----|----|
| $m$ | -5.51 | -5.42 | -5.46 | -5.42 | -5.41 | -5.41 | -5.46 | -5.42 |
| $s$ | 0.0021 | 0.0084 | 0.015 | 0.011 | 0.0086 | 0.0093 | 0.015 | 0.012 |
| $m/n_{at}$ | -8.74 | -8.60 | -8.67 | -8.60 | -8.59 | -8.59 | -8.67 | -8.60 |
| $s/n_{at}$ | 0.0033 | 0.013 | 0.024 | 0.018 | 0.014 | 0.015 | 0.024 | 0.018 |

As above, but for an AFeFPc molecule. Shaded cells indicate adsorption sites for which the corrugation of the surface-molecule interaction potential (as measured by $s/n_{at}$) decreased relative to the case of FeFPc.

| Site | 1 | 3 | 6 | 8 | 11 | 13 | 16 | 18 |
|---|---|---|---|---|---|---|---|---|
| $m$ | -5.14 | -5.00 | -5.08 | 5.00 | -5.00 | -5.00 | -5.07 | -5.00 |
| $s$ | 0.00050 | 0.051 | 0.022 | 0.027 | 0.051 | 0.051 | 0.022 | 0.025 |
| $m/n_{at}$ | -9.02 | -8.77 | -8.90 | -8.77 | -8.77 | -8.77 | -8.90 | -8.77 |
| $s/n_{at}$ | 0.00088 | 0.089 | 0.038 | 0.047 | 0.089 | 0.089 | 0.038 | 0.043 |

As above, but for an MnHPc molecule.

| Site | 1 | 3 | 6 | 8 | 11 | 13 | 16 | 18 |
|---|---|---|---|---|---|---|---|---|
| $m$ | -5.60 | -5.47 | -5.54 | -5.46 | -5.46 | -5.46 | -5.54 | -5.48 |
| $s$ | 0.0028 | 0.051 | 0.029 | 0.031 | 0.05 | 0.05 | 0.029 | 0.028 |
| $m/n_{at}$ | -8.89 | -8.68 | -8.79 | -8.68 | -8.67 | -8.67 | -8.79 | -8.68 |
| $s/n_{at}$ | 0.0045 | 0.081 | 0.046 | 0.049 | 0.079 | 0.081 | 0.046 | 0.046 |

As above, but for an AMnHPc molecule. Shaded cells indicate adsorption sites for which the corrugation of the surface-molecule interaction potential (as measured by $s/n_{at}$) decreased compared to the case of MnHPc.

**Supporting information 2. Comparison of GPR energy predictions with DFT-calculated energies**

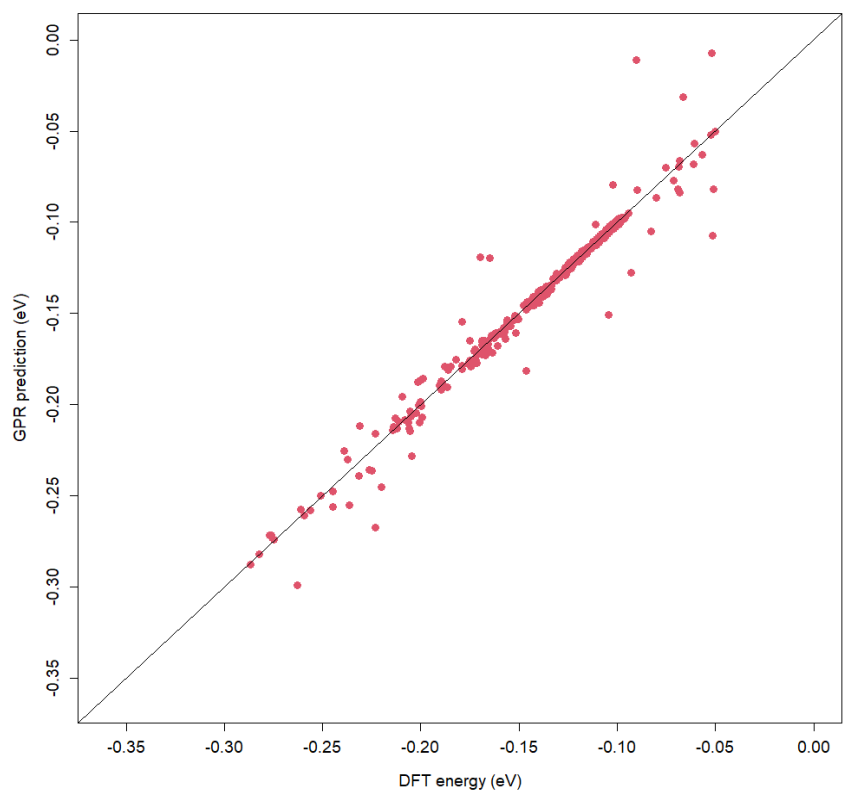

Comparison for the symmetric molecule case using 695 randomly generated pairwise interactions as test data.

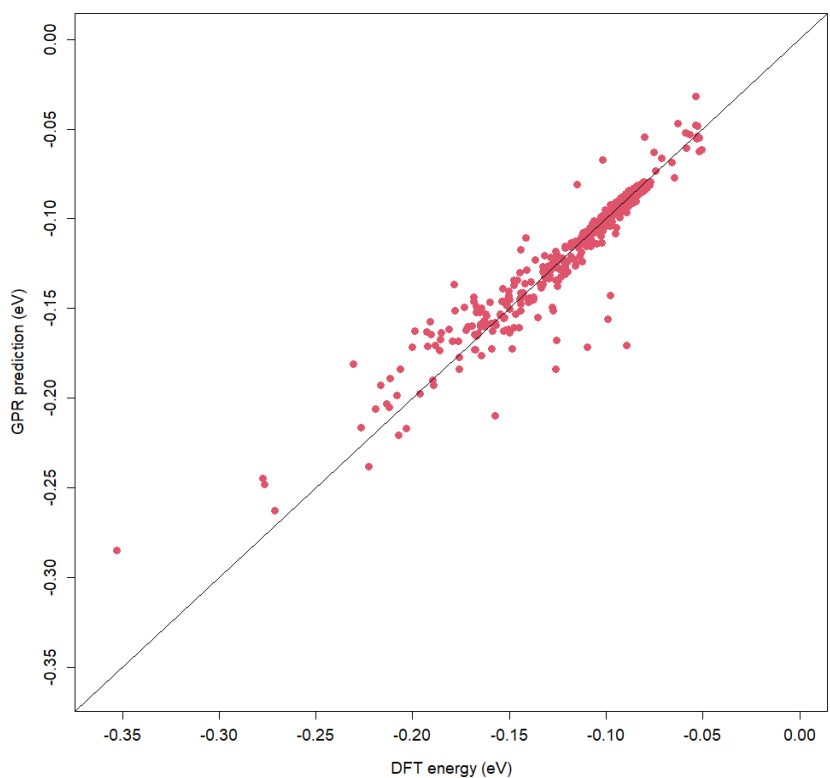

Comparison for the asymmetric molecule case using 696 randomly generated pairwise interactions as test data.

# Supporting Information 3: Formula for the configuration weight

Consider the configuration $c_i$ in the top-left side of Figure 3A. This configuration contains two groups of molecules, as indicated by the blue dotted lines. The $W$ in equation (5) is the number of ways that these two groups can be simultaneously placed on the surface in arbitrary orientations. According to this definition, $W$ is infinitely large for the case of an infinitely large surface. However, in the EUF method, the entropy in equation (4) is not a 'physical' entropy and does not need to be treated rigorously. The entropy term is only used to facilitate the configuration space search, and providing that the pseudotemperature $\tau$ is approximately zero by the end of the EUF run, it will make a negligible contribution to the free energy of the final result. We can therefore choose any formula that we wish for $W$, providing that it improves the success probability of EUF.

In this work, $W$ is given by a formula which makes three assumptions. Namely, (1) a *finite* number of adsorption sites, (2) that each group of molecules occupies an infinitesimally small area of the surface, and (3) that there are no interactions between different groups of molecules. Assumption (1) implies that $W$ is finite. Assumption (2) implies that each group of molecules occupies only one adsorption site. Assumption (3) means that in computing $W$, we can place the groups of molecules as close together as we like without worrying about interactions between them. This formula was studied in detail in previous work (see reference [39] of the main text).

Consider a finite grid of $N$ adsorption sites upon which $m$ groups of molecules reside. Suppose that these $m$ groups can be further partitioned into $M$ supersets, such that two groups of molecules belong to the same superset if and only if they can be perfectly superimposed by arbitrary translations and rotations. Under the three assumptions above, the number of ways in which these $m$ groups can be placed on the surface is

$$W = \frac{N!}{(N-M)!h_1!h_2!\cdots h_M!} \times R_1 R_2 \cdots R_M,$$

where $h_k$ is the size of superset $k$ and $R_k$ is the number of ways in which orientations can be assigned to the groups in superset $k$. The left-hand factor is the number of ways in which the $N$ adsorption sites can be partitioned into $N - M$ empty sites, $h_1$ sites for the groups of superset 1, $h_2$ sites for the groups of superset 2, and so on. $R_k$ is defined as

$$R_k = n_o + \sum_{\{h_{k1},h_{k2}:h_{k1}+h_{k2}=h_k\}} \frac{h_k!}{h_{k1}!h_{k2}!} n_o(n_o-1)$$
$$+ \sum_{\{h_{k1},h_{k2},h_{k3}:h_{k1}+h_{k2}+h_{k3}=h_k\}} \frac{h_k!}{h_{k1}!h_{k2}!h_{k3}!} n_o(n_o-1)(n_o-2)$$
$$+ \cdots$$
$$+ h_k! n_o(n_o-1)(n_o-2)\cdots(n_o-h_k),$$

where $n_o$ is the number of orientations available to the for the molecules. The first term (the first $n_o$ after the equals sign) is for the case where all groups of molecules are assigned the same orientation. The second term is for the case where $h_{k1}$ groups of molecules have one orientation and $h_{k2}$ groups have a different orientation. And so on. Note that this formula assumes that $n_o$ is greater than $h_k$. If $h_k$ is greater than $n_o$, then the sum terminates at the last non-negative term.

In addition to the pseudotemperature $\tau$, the above formula introduces a second parameter $N$. Like $\tau$, we are free to choose any (positive, interger) value for this parameter in order to adjust the performance of EUF. In this work, $N$ was set to 2500.

**Supporting Information 4: EUF+MCMC compared to ordinary MCMC with a random initial condition**

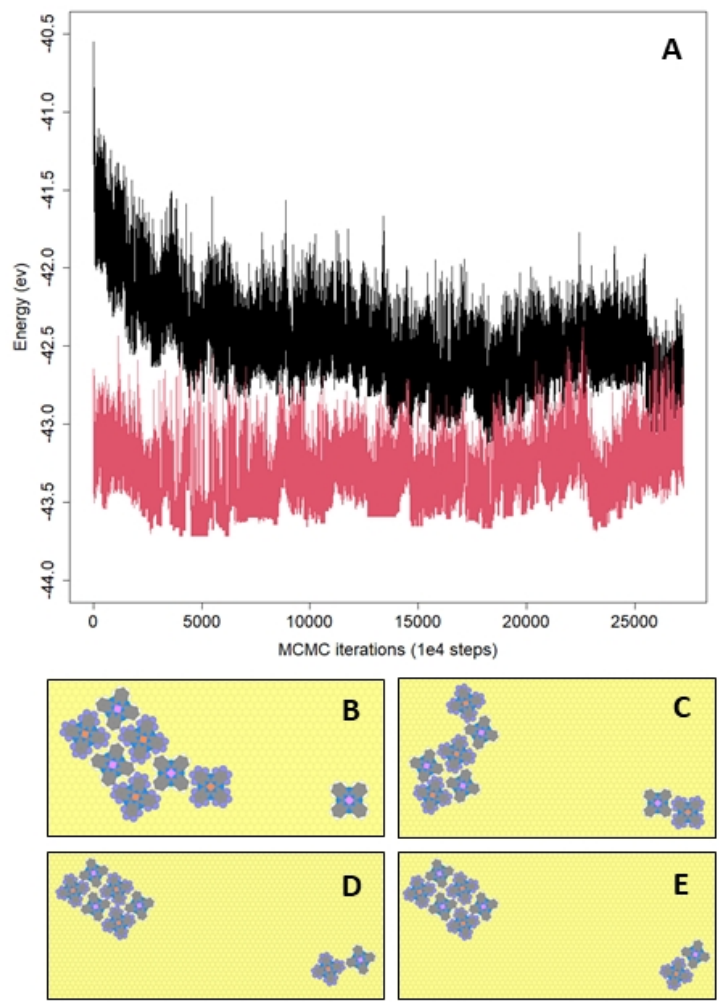

(A) Performance of ordinary MCMC with a random initial condition (black curve) compared to MCMC with EUF-generated initial condition (EUF+MCMC, red curve). "Energy" refers to the energy of the sampled configuration. (B – E) Four typical configurations sampled during the second half of the ordinary MCMC trajectory.

**Supporting Information 5: Comparison of configuration $s_1$ with experimental data**

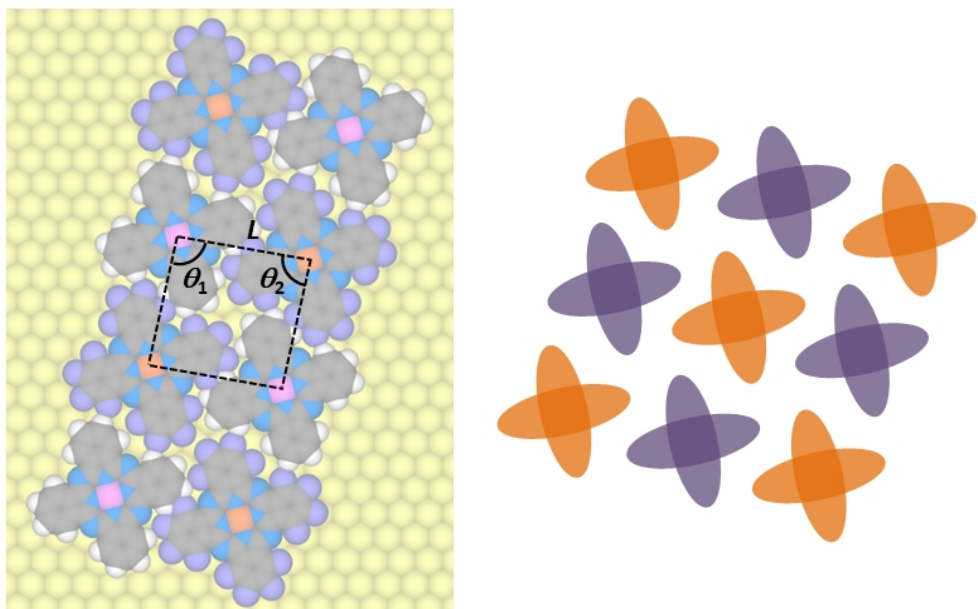

Configuration $s_1$ (left) and schematic reproduction of a scanning tunneling microscopy (STM) image (right). The image on the right is based on Figure 1 of Girovsky *et al. Nat. Commun.* **8**, 2017, 15388.

|  | Calculation (mean ± standard error) | Experiment (mean ± standard error) |
|---|---|---|
| $L$ | 14.14 ± 0.07 | 14.05 ± 0.08 |
| $\theta_1$ | 95.27 ± 1.87 | 95.84 ± 0.26 |
| $\theta_2$ | 84.73 ± 1.87 | 84.84 ± 0.79 |

Comparison of Fe-Mn distances ($L$) and angles between metal ions ($\theta_1$ and $\theta_2$) between configuration $s_1$ (left column) with those measured from Figure 1 of Girovsky *et al* (right column). Experimental measurements were obtained by directly analysing the figure using the ImageJ software (ImageJ 1.52V, http://imagej.nih.gov/ij). For the experimental image, $L$, $\theta_1$, and $\theta_2$ were estimated from 12, 4, and 4 measurements, respectively.

**Supporting Information 6: Lowest energy pairwise interactions from random sample**

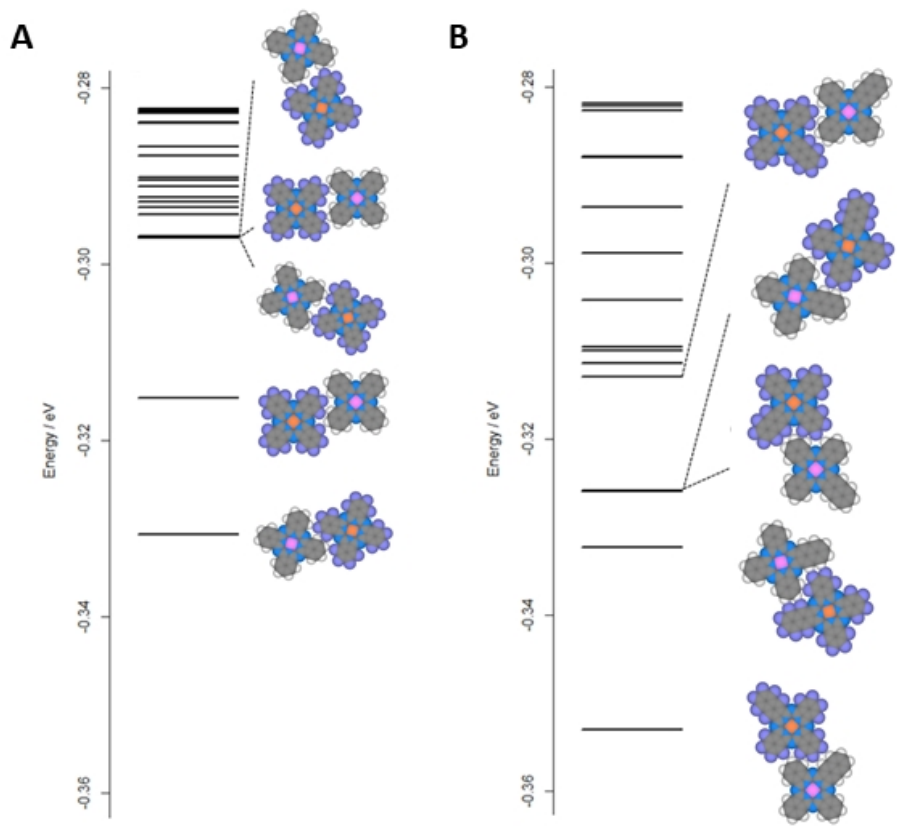

Distribution of energies from a sample of 9000 randomly generated pairwise interactions. (A) is for the case of symmetric molecules, and (B) is for the case of asymmetric molecules. Only the low-energy tail of the distribution (energy < -0.28 eV) is shown. Atomistic models of the five lowest energy interactions are shown in both cases.

**Supporting Information 7: Magnetic gaps for 300 K configuration sample computed without screening of RKKY interactions**

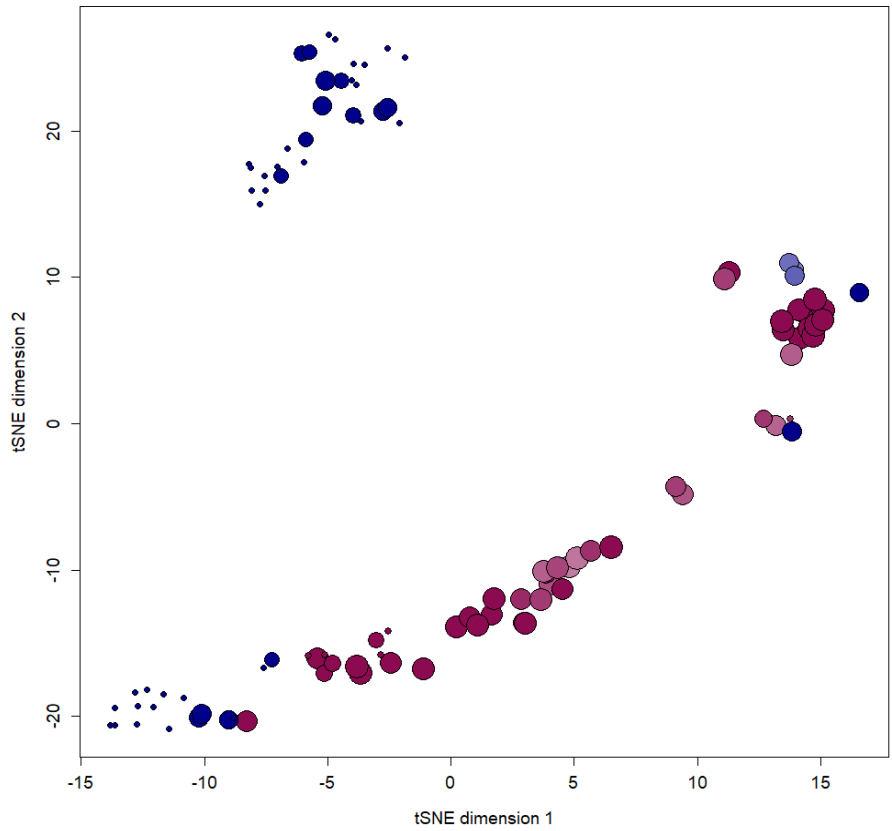

The points are coloured according to the key in Figure 4A. Here, the magnetic gaps were calculated according to the same model as in equations (7- 10) but with screening factor $\sigma_{ij}$ set to 1.